\documentclass[prd,amsmath,amsfonts,showpacs,
               superscriptaddress,nofootinbib,showkeys,floatfix,%
               twocolumn,a4paper]{revtex4}

\usepackage{amssymb,amsmath,amsfonts}
\usepackage{graphicx}
\usepackage{ifpdf}
\usepackage[hyperfootnotes=false]{hyperref}
\newcommand{\Eq}[1]{Eq.~(\ref{#1})}

\newcommand{\Fig}[1]{Fig.~\ref{#1}}
\newcommand{\Figs}[1]{Figs.~\ref{#1}}

\newcommand{\ndf}{\mathsf{ndf}}                     

\renewcommand{\max}{\mathrm{max}\mbox{-}}

\def\eq#1{(\ref{#1})}
\def\Eq#1{Eq.~(\ref{#1})}

\ifpdf
\graphicspath{{./Fig-PDF/}}
\else
\graphicspath{{./Fig/}}
\fi

\begin{document}

\preprint{}

\newcommand{\AU}{Centre for the Subatomic Structure of Matter, 
  University of Adelaide, SA 5005, Australia}
\newcommand{\GU}{Institut f\"ur Physik, Karl-Franzens Universit\"at
  Graz, Universit\"atsplatz 5, A-8010 Graz, Austria}  
\newcommand{\HU}{Institut f\"ur Theoretische Physik, Universit\"at
  Heidelberg, Philosophenweg 16, D-69120 Heidelberg, Germany} 
\newcommand{\TUD}{Institut f\"ur Kernphysik, Technische Universit\"at
  Darmstadt, Schlossgartenstr.~9, D-64289 Darmstadt, Germany}
\newcommand{\EMMI}{ExtreMe Matter Institute EMMI, GSI Helmholtzzentrum f\"ur 
Schwerionenforschung, Planckstr.~1, D-64291 Darmstadt, Germany}
\newcommand{\RGB}{Institut f\"ur Theoretische Physik, Universit\"at Regensburg,
  D-93040 Regensburg, Germany}

 \title{Strong-coupling study of the Gribov ambiguity in lattice Landau gauge} 

 \author{Axel Maas}
 \affiliation{\GU}

 \author{Jan M.~Pawlowski}
 \affiliation{\HU}
\affiliation{\EMMI}
 \author{Daniel Spielmann}
 \affiliation{\HU}
 \affiliation{\EMMI}
 \author{Andr\'e Sternbeck}
 \affiliation{\AU}
 \affiliation{\RGB}

 \author{Lorenz von~Smekal}
 \affiliation{\TUD}

\begin{abstract}
  We study the strong-coupling limit $\beta=0$ of lattice SU(2) Landau
  gauge Yang--Mills theory. In this limit the lattice spacing is
  infinite, and thus all momenta in physical units are infinitesimally
  small. Hence, the infrared behavior can be assessed at sufficiently
  large lattice momenta. Our results show that at the lattice volumes
  used here, the Gribov ambiguity has an enormous effect on the ghost
  propagator in all dimensions. This underlines the severity of the
  Gribov problem and calls for refined studies also at finite
  $\beta$. In turn, the gluon propagator only mildly depends on the
  Gribov ambiguity.
\end{abstract}



\maketitle

\section{Introduction}
\label{sec:intro}

The study of the long-range behavior of QCD Green's functions is one
path towards the understanding of confinement
\cite{Alkofer:2000wg,Alkofer:2006fu}. In particular, the confining
properties of QCD are directly reflected in the behavior of the Landau
gauge gluon and ghost propagators at asymptotically low momenta
\cite{Alkofer:2003jj}. Furthermore, confinement, as signaled by an
unbroken center symmetry, can also be shown to be a direct consequence
of these properties \cite{Braun:2007bx}.

However, implementing Landau gauge as in the perturbative regime is
not sufficient to uniquely fix the infrared behavior of Green's
functions because of the Gribov--Singer ambiguity
\cite{Gribov:1977wm,Singer:1978dk}. Using functional methods, one
finds a one-parameter family of solutions which can be classified by
the zero momentum value of the ghost propagator
\cite{Fischer:2008uz,Boucaud:2008ky}. These can be obtained by
specifying a non-perturbative renormalization
condition~\cite{Fischer:2008uz}.

The one-parameter family of solutions contains the {\em scaling
  solution} as a limiting special case \cite{Fischer:2008uz}, which is
characterized by an infrared divergent ghost dressing function, and an
infrared suppressed and likely vanishing gluon propagator, see e.g.\
\cite{vonSmekal:1997is,vonSmekal:1997vx,Alkofer:2000wg,%
  Zwanziger:2001kw,Lerche:2002ep,Pawlowski:2003hq,%
  Alkofer:2004it,Fischer:2006vf,Huber:2007kc,%
  Zwanziger:2009je}. In contrast, the other members of the family
with an infrared finite ghost dressing function necessarily come along
with an infrared finite gluon propagator, see e.g.\
\cite{Aguilar:2004sw,Aguilar:2008fh,%
  Aguilar:2008xm,Boucaud:2006if,Boucaud:2008ji,Boucaud:2008ky,%
  Dudal:2007cw,Dudal:2008sp,Fischer:2008uz,Binosi:2009qm,%
  Aguilar:2009nf}. These are called the {\em decoupling solutions}
\cite{Fischer:2008uz}. Both terms are more precisely defined in
Section \ref{sec:ISvD}.

These Green's functions have also been determined on the lattice. By
now it is well established that standard lattice gauge fixings lead to
a decoupling-type behavior, see e.g.\
\cite{Bakeev:2003rr,Silva:2004bv,Bogolubsky:2005wf,Sternbeck:2005tk,%
  Bogolubsky:2007bw,Cucchieri:2007rg,Sternbeck:2007ug,Cucchieri:2008fc,
  Bornyakov:2008yx,Bogolubsky:2009dc,Pawlowski:2009iv}, though not unanimously \cite{Oliveira:2009eh,Oliveira:2009nn}. Moreover,
investigations of the Gribov--Singer ambiguity in this setting,
\cite{Cucchieri:1997dx,Cucchieri:1997ns,Bakeev:2003rr,Silva:2004bv,%
  Sternbeck:2005tk,Bogolubsky:2005wf,Bornyakov:2008yx,%
  Bogolubsky:2007bw,Sternbeck:2008wg,Sternbeck:2008mv,Sternbeck:2008na,%
  Maas:2008ri,Bogolubsky:2009dc,Pawlowski:2009iv}, so far only led to
relatively small, quantitative effects.

However, recently alternative gauge fixing procedures have been
devised that directly make use of the Gribov--Singer ambiguity
\cite{Maas:2009se}. Here, one selects different non-perturbative
completions of the Landau gauge by constricting the ghost propagator
in the zero momentum limit to be as close as possible to a fixed
value. Such a procedure samples Gribov copies differently than a
standard gauge fixing. It introduces an additional parameter which
plays a role analogous to the non-perturbative renormalization
condition of the continuum studies. The interpretation here is that
this free parameter is therefore linked to the residual gauge freedom
of the Landau gauge.

Note that the qualitatively different infrared solutions, decoupling
and scaling, signal different realizations of the global part of the
gauge fixing. At the level of gauge-invariant observables they have to
be equivalent.  One particular difference is the structure of the
Becchi--Rouet--Stora--Tyutin (BRST) symmetry. If a given non-perturbative
completion of the Landau gauge is to have a global BRST symmetry then
the Kugo--Ojima confinement criterion requires that the corresponding
Landau gauge ghost propagator is infrared enhanced, corresponding to
the scaling solution \cite{Fischer:2008uz,PawlowskiSmekal}. On the
other hand, in the decoupling case, the infrared ghost propagator is
essentially massless. In standard Faddeev--Popov theory, the
corresponding massless asymptotic ghost states, albeit being
unphysical, would then contribute to the global gauge charges and lead
to a spontaneous breaking of the global gauge symmetry much like what
one expects for theories with a Higgs mechanism. When there is no
Higgs mechanism, and thus no physical vector field in the gauge boson
sector, this breaking is most likely artificial and due to a
non-perturbative breakdown of BRST symmetry itself
\cite{Dudal:2007cw,Dudal:2008sp,Tissier:2008nw,Tissier:2009sm,%
  Gomez:2009tj,Sorella:2009vt,Kondo:2009wk,Kondo:2009gc,%
  Kondo:2009qz,Kondo:2009ug}.  At present, the only solution which is
consistent with an unbroken BRST symmetry is thus the scaling
solution. Note that other BRST formulations might exist, in which the
massless ghosts of the decoupling solutions would not automatically
lead to BRST breaking. However, such a formulation is still lacking,
but investigations along the lines of \cite{Sorella:2009vt} appear
promising.

On the lattice, the only way that is currently known to define BRST
symmetry in presence of Gribov copies and thus fully
non-perturbatively is based on defining gauge fields by stereographic
projection. This avoids the perfect cancelation of Gribov copies and
the Neuberger 0/0 problem of lattice BRST, while it nevertheless
reduces to the standard BRST symmetry in the continuum limit,
\cite{Neuberger:1986xz,vonSmekal:2008ws,%
  vonSmekal:2008es,vonSmekal:2007ns,Mehta:2009zv}. 

As noted above, for studying the long-range behavior of the Landau
gauge gluon and ghost propagators on the lattice, it is useful to
employ the strong-coupling limit
\cite{Sternbeck:2008wg,Sternbeck:2008mv,Sternbeck:2008na,Cucchieri:2009zt,%
  Cucchieri:1997fy}. In our present work we use the strong-coupling
limit $\beta \to 0$ for pure SU(2) lattice gauge theory in $d=2$ and
$3$ dimensions.  This limit can be interpreted as the limit of
infinite lattice spacing, $a \to \infty$, at a fixed physical scale as
set, e.g., by the string tension. Alternatively, when considering all
momenta in lattice units, $q \sim 1/a$, the same limit can also be
interpreted as the hypothetical limit in which the physical scale is
sent to infinity, i.e., formally as
$\Lambda_\mathrm{QCD}\to\infty$. Both interpretations are
equivalent. They are based on taking the limit of the function
$a(\beta)$ for $\beta\to 0$. 
Other ways to assign a scale to the theory at
$\beta = 0$ are possible \cite{deForcrand:2009dh,Cucchieri:2009zt},
but they require different prescriptions. For the limit $\beta\to 0$
adopted here, all momenta and masses in lattice units $1/a$ are
infinitely small relative to the physical scale of the theory which is
precisely what we need for an analysis of the asymptotic infrared
behavior of its correlation functions. It is therefore best 
suited to isolate this behavior from finite-volume effects. 

In recent studies
\cite{Sternbeck:2008wg,Sternbeck:2008mv,Sternbeck:2008na,Cucchieri:2009zt}
the $\beta \to 0$ limit was first investigated only in four dimensions
\cite{Sternbeck:2008wg,Sternbeck:2008mv,Sternbeck:2008na} and then to
some extent also in lower dimensions \cite{Cucchieri:2009zt}. The
results of both studies were compatible with a scaling behavior for
the gluon propagator for intermediate to large momenta and a
decoupling behavior at small momenta, where `large' and `small' in the
strong-coupling limit always refers to lattice units. It has moreover
been shown in
\cite{Sternbeck:2008wg,Sternbeck:2008mv,Sternbeck:2008na} that the
mass parameter related to the low-momentum decoupling tail can be
changed by choosing different definitions of the gluon field, which
should be equivalent in the continuum limit. For non-vanishing lattice
spacing $a$, however, these correspond to different lattice
implementations of the Landau gauge \cite{vonSmekal:2008es}. If some
of these differences remain in the continuum limit that would
strengthen the hypothesis
\cite{Fischer:2008uz,Maas:2009se,Maas:2008ri} of a residual gauge
freedom related to the one-parameter family of solutions observed in
the continuum. In any case, the lattice discretization dependence adds
to the puzzle of the global properties of Landau gauge Yang--Mills
theory.

The interpretation of the data for the ghost propagator has always
been even less conclusive, and did not allow a direct determination of
local scaling coefficients.  This led to a more indirect
interpretation of the data. In
\cite{Sternbeck:2008wg,Sternbeck:2008mv,Sternbeck:2008na} it was
concluded that the ghost data at larger lattice momenta could be
consistent with the scaling behavior observed in the gluon
propagator. However, it was not possible to extract a firm value for
the scaling exponent from the ghost data for any reasonably wide range
of lattice momenta. In turn, the authors of \cite{Cucchieri:2009zt}
assessed the evidence of a logarithmic momentum dependence and argued
in favor of decoupling at all lattice momenta for Landau gauge
Yang--Mills theory in three and four dimensions in the strong-coupling
limit. It was also frequently claimed that the scaling solution is
realized in two dimensions in agreement with the arguments in
\cite{Dudal:2008xd} and the lattice data of
\cite{Maas:2007uv,Cucchieri:2007rg,Pawlowski:2009iv}.

In this paper we present results with much improved statistical
accuracy in two and three dimensions. We also provide a more detailed
study of the systematic uncertainties, see Appendices \ref{app:source}
and \ref{app:momenta}, and of the relevance of the Gribov ambiguity,
see Section \ref{sec:landauB} and parts of Sec.~\ref{sec:results}.

Our results are presented in Sec.~\ref{sec:results}. For the gluon
propagator we confirm the scaling results of
\cite{Sternbeck:2008wg,Sternbeck:2008mv,Sternbeck:2008na,%
  Cucchieri:2009zt} at intermediate lattice momenta. In particular, we
obtain fairly stable scaling exponents also with the local analysis of
\cite{Cucchieri:2009zt}. The values for the exponents tend to decrease
somewhat at large lattice momenta, and as in the previous studies we
observe decoupling at small momenta, at least in three dimensions.

A qualitatively new and perhaps surprising behavior is seen in the
ghost propagator which shows a very strong sensitivity to Gribov
copies at low momenta. This entails that we still have to work on a
better understanding of the global properties of Landau gauge QCD. In
particular our analysis shows that a proper treatment of Gribov copies
in the thermodynamical limit of Landau gauge lattice Yang--Mills theory
has not been achieved yet. Subject to the sampling of Gribov copies
the ghost propagator can be tuned to also show a scaling behavior for
intermediate momenta. The general trend is that even a drastic
over-scaling is obtained if enough gauge copies are taken into account
in the search for the maximally achievable infrared enhancement. In
fact, the possibility of such an over-scaling has also been observed in
continuum Landau gauge Yang--Mills theory on the torus
\cite{Fischer:2005ui}. Here, we moreover demonstrate that this
peculiar behavior arises already in two dimensions where infrared
scaling was supposed to be quite well established. We find that the
two-dimensional ghost propagator shows the same apparently
unconstrained infrared behavior as that in higher dimensions, and that
there is thus no qualitative difference between two and higher
dimensions as long as sufficiently many Gribov copies are present, as
has been suspected \cite{Maas:2008ri,Maas:2009se}.

In the summary in Section \ref{sec:conclusion} we are led to conclude
that our analysis clearly shows the direct relation between the
ambiguities in the ghost propagator at low and intermediate momenta
and the Gribov ambiguity. In our opinion this ambiguity is much more
than an artificial problem of this unphysical limit. In fact, Gribov
copies at finite lattice coupling $\beta$ will remain to affect the
infrared behavior of Green's functions as long as they are present in
the strong-coupling limit. To resolve it there will certainly be more
cost-efficient than to try to reach the continuum limit in a
sufficiently large physical volume by brute force, though this is
attempted as well \cite{Maas:prep}.

\section{Infrared Scaling and Decoupling}
\label{sec:ISvD}

The Landau gauge gluon propagator, in (Euclidean) momentum space, is
parameterized by a single dressing function $Z$,
\begin{equation}
  \label{eq:gluonprop_dress}
  D^{ab}_{\mu\nu}(p) \,=\, 
  \delta^{ab}\left(\delta_{\mu\nu}-\frac{p_{\mu}p_{\nu}}{p^2}\right) 
  \frac{Z(p^2)}{p^2} \; ,
\end{equation}
and the ghost propagator by a corresponding dressing function $G$,
\begin{equation}
 \label{eq:ghostprop_dress}
  D_G^{ab}(p)   \,=\,  -\delta^{ab}\;\frac{G(p^2)}{p^2} \; .
\end{equation}
For their infrared behavior, i.e., that of $Z(p^2)$ and $G(p^2) $
for $p^2 \to 0$, we consider the two possibilities of scaling and
decoupling.  Occasionally, we also use $D_{gl}=Z(p^2)/p^2$ and
$D_{gh}=G(p)^2/p^2$.

\subsection{Scaling}

The prediction of
\cite{vonSmekal:1997is,vonSmekal:1997vx,Alkofer:2000mz,Lerche:2002ep,
  Zwanziger:2001kw,Pawlowski:2003hq} amounts to infrared asymptotic
forms
\begin{subequations}
   \label{eq:infrared-gh_gl}
\begin{align}
  \centering
  \label{eq:infrared-gl}
  Z(p^2) \, &\sim\, (p^2/\Lambda^2_\mathrm{QCD})^{2\kappa_Z+\frac{4-d}{2}} \; ,\\
 \label{eq:infrared-gh}
  G(p^2) \, &\sim \, 
  (p^2/\Lambda_\mathrm{QCD}^2)^{-\kappa} \; ,
\end{align}
\end{subequations}
for $p^2 \to 0$, which are both determined by an unique critical
infrared exponent
\begin{equation}
  \label{kappaZ=kappaG}
   \kappa_Z= \kappa \; , 
\end{equation}
with $ (d-2)/4 \le \kappa < d/4$. Furthermore it has been proven in
\cite{Braun:2007bx} that the scaling solution is confining for 
\begin{equation}\label{eq:confcrit}
\kappa > \frac{d-3}{4}\,.
\end{equation}
Under a mild regularity assumption
on the ghost--gluon vertex \cite{Lerche:2002ep}, the value of this
exponent is furthermore obtained in four dimensions as
\cite{Lerche:2002ep,Zwanziger:2001kw}   
\begin{equation} 
  \kappa \, = \, (93 - \sqrt{1201})/98 \, \approx \, 0.595 \; , 
\label{kappa_c}
\end{equation}
which is confinement according to \eq{eq:confcrit}. The corresponding
values in lower dimensions are $\kappa = 1/5$ ($d=2$) and $\kappa
\approx 0.3976$ ($d=3$) which also satisfy \eq{eq:confcrit}. The
conformal nature of this infrared behavior in the pure Yang--Mills
sector of Landau gauge QCD is evident in the generalization to
arbitrary gluonic correlations \cite{Alkofer:2004it}: an uniform
infrared limit of one-particle irreducible vertex functions
$\Gamma^{m,n}$ with $m$ external gluon legs and $n$ pairs of
ghost/anti-ghost legs, in four dimensions of the form
\begin{equation}
  \Gamma^{m,n} \, \sim\, (p^2/\Lambda_\mathrm{QCD}^2)^{(n-m)\kappa}\; ,   
\label{genIR}
\end{equation}
when all $p_i^2 \propto p^2 \to 0$, $i=1,\dots, 2n+m$. In particular,
the ghost--gluon vertex is then infrared finite (with $n=m=1$), in
agreement with its STI \cite{Taylor:1971ff}, and the non-perturbative
running coupling introduced in
\cite{vonSmekal:1997is,vonSmekal:1997vx} via the definition
\begin{equation} 
\alpha_s(p^2) \, = \, \frac{g^2}{4\pi} Z(p^2) G^2(p^2) 
\label{alpha_minimom}
\end{equation}
approaches an infrared fixed point, $\alpha_s \to \alpha_c$ for $p^2
\to 0$ in $d=4$. For arbitrary dimensions, trivial kinematic factors
lead to an infrared fixed point for the effective coupling
\begin{equation}
\alpha_e(p^2)\,=\, p^{d-4}\alpha_s(p^2) \label{eq:def-alphas},
\end{equation}
with $\alpha_e(0)=\alpha_c$.
If the ghost--gluon vertex is regular at $p^2 =0$, its value is in four dimensions
\cite{Lerche:2002ep}
\begin{equation} 
\alpha_c \, = \, \frac{8\pi}{N_c} \, \frac{\Gamma^2(\kappa-1)
  \Gamma(4-2\kappa)}{\Gamma^2(-\kappa) \Gamma(2\kappa-1)} \,  \approx
  \, \frac{9}{N_c} \times \, 0.99  \; . 
  \label{eq:alphac}
\end{equation}
Comparing the infrared scaling behavior of Dyson-Schwinger equation
(DSE) and functional renormalization group equation (FRGE) for
solutions of the form of Eqs.~(\ref{eq:infrared-gh_gl}), it has been
shown that in presence of a single scale, the QCD scale
$\Lambda_\mathrm{QCD}$, the solution with the infrared behavior
(\ref{kappaZ=kappaG}) and (\ref{genIR}), with positive $\kappa$, is
unique
\cite{Fischer:2006vf,Fischer:2009tn,Huber:2007kc,Huber:2009tx}. It is
nowadays being called the \emph{scaling solution}.

\subsection{Decoupling}

This uniqueness proof does not rule out, however, the possibility of a
solution with an infrared finite gluon propagator, as arising from a
transverse gluon mass $M$, which then leads to an essentially free
ghost propagator, with the free massless-particle singularity at
$p^2=0$, i.e.,
\begin{equation} 
  Z(p^2) \, \sim \, p^2/M^2\,, \quad \text{and} \quad G(p^2) \, \sim \,
  \mathrm{const.}
  \label{decoupling}
\end{equation}
for $p^2 \to 0$ \cite{Aguilar:2004sw,Aguilar:2008fh,%
  Aguilar:2008xm,Boucaud:2006if,Boucaud:2008ji,Boucaud:2008ky,%
  Dudal:2007cw,Dudal:2008sp,Fischer:2008uz,Binosi:2009qm,%
  Aguilar:2009nf}. The constant contribution to the zero-mo\-men\-tum
gluon propagator, $ D(0) = 1/M^2$, thereby necessarily leads to an
infrared constant ghost renormalization function $G$. This solution
corresponds to $\kappa_Z = 1/2 $ and $ \kappa = 0$, and  it has been proven
in \cite{Braun:2007bx} that it is confining. It does not satisfy the
scaling relations (\ref{kappaZ=kappaG}) or (\ref{genIR}).  This is
since the transverse gluons decouple for momenta $p^2 \ll M^2 $, below
the independent second scale given by their mass $M$. It is thus not
within the class of scaling solutions considered above, and it is
termed the {\em decoupling solution} in contradistinction. The
renormalization group invariant (\ref{alpha_minimom}) shows this
decoupling as it tends to zero in the infrared. Alternatively, we can
define a running coupling analogously to that in massive theories. By
fixing one overall free parameter this definition leads to a coupling
which even quantitatively resembles the fixed-point coupling in the
scaling case. This definition has been given explicitly in
\cite{Fischer:2008uz} and similarly been used since then
\cite{Aguilar:2009nf,Binosi:2009qm}.

\section{Gauge fixing and the ghost propagator}
\label{sec:landauB}

The functional equations of continuum quantum field theory, such as
DSEs and FRGEs, admit a one-parameter family of solutions
\cite{Fischer:2008uz}. The free parameter can be chosen as the value
of the ghost dressing function at vanishing momentum.  It fixes an
ambiguity in the equations related to non-perturbative renormalization
and the structure of the massless unphysical states in the theory
which is a priori unknown. The realization of the global gauge
symmetries and the Kugo--Ojima confinement criterion crucially depend
on this structure. For that reason functional equations have been
supplemented by the appropriate boundary condition in the earlier
literature on the scaling solution \cite{Lerche:2002ep}. The
restriction of the gauge field configuration space to the first Gribov
region provides a simple example of boundary conditions which are not
reflected in the form of the functional equations of the theory but
need to be imposed in addition \cite{Zwanziger:2001kw}. Another
example is the Kugo--Ojima criterion which cannot be derived from the
functional equations but needs to be added as a boundary condition.
This then singles out the scaling solution with an infrared
enhancement of the ghost propagator. Without this boundary condition,
the functional equations admit both, the scaling solution and the
decoupling solutions. Since these boundary conditions are imposed on
unphysical degrees of freedom, they may be inconsequential for
physical observables.

Indeed, the Gribov ambiguity and the associated freedom in globally
completing the Landau gauge are well known to affect particularly the
ghost propagator on the lattice
\cite{Cucchieri:1997dx,Cucchieri:1997ns,Bakeev:2003rr,Silva:2004bv,%
  Sternbeck:2005tk,Bogolubsky:2005wf,Bornyakov:2008yx,%
  Bogolubsky:2007bw,Sternbeck:2008wg,Sternbeck:2008mv,Sternbeck:2008na,%
  Maas:2008ri,Maas:2009se,Bogolubsky:2009dc,Pawlowski:2009iv}. For the volumes
accessible in current simulations, it appears that this ambiguity is
indeed closely related to the one-parameter freedom in the family of
solutions to the functional equations: a direct correspondence between
the Gribov ambiguity on the lattice and the one-parameter family of
solutions to the functional equations of continuum quantum field
theory is provided by the so-called Landau-$B$ gauges
\cite{Maas:2009se}. These gauges correspond to different
resolutions of the Gribov--Singer ambiguity. They are implemented by
first selecting a target value, called the $B$ parameter, for the
ratio of ghost dressing functions evaluated at two widely separated
momentum scales. An appropriate choice for the two momentum scales is
the lowest accessible momentum on the lattice and some conveniently
large momentum scale, such as the renormalization scale. This choice
is made to resemble the non-perturbative freedom in choosing a
boundary condition for the ghost propagator in the continuum
studies. Then, on a configuration by configuration basis, many Gribov
copies are generated to select the one where this ratio is closest to
the target value $B$. The propagators, or other Green's functions, are
calculated on these copies from the Monte-Carlo history of
configurations.

Any finite number of Gribov copies leads to a finite set of proposed
$B$ values. In turn, for larger and larger lattices one can get close
to any target value within a certain range depending on the lattice
volume. We conclude that within this range the infrared value of the
ghost dressing function can be chosen freely. An extreme case is the
$\max B$ gauge \cite{Maas:2009se} in which the copy with the maximally
enhanced ghost propagator is selected.

In general, it is of course practically impossible to find all Gribov
copies for a given configuration. This limitation can cause some
residual Gribov noise on top of the statistical fluctuations. In the
case of the $\max B$ gauge the result will be a lower bound for the
maximally enhanced ghost propagator that is possible in any finite
volume. In the present work, $\max B$ gauge is implemented similarly
to \cite{Maas:2009se}. In order to reduce the effects of the lattice
breaking of rotational invariance, the Euclidean $O(d)$ symmetry, the
value of $B$ for a given copy is determined by averaging over all
representations for the lowest momentum, instead of using only one
on-axis momentum as in \cite{Maas:2009se}.  Note also that the
standard lattice implementation of the Landau gauge, the minimal
Landau gauge, essentially gives the same results as choosing for $B$
the expectation value of the ghost propagator ratio from all Gribov
copies of all configurations \cite{Maas:2009se}.

\section{Numerical results}
\label{sec:results}

The configurations have been obtained by generating randomly $\beta=0$
configurations on symmetric lattices of size
$L^d=(Na)^d$. Gauge fixing to the minimal Landau gauge in sections
\ref{subsec:results2d} and \ref{subsec:results3d} has been performed 
using a stochastic overrelaxation method, see
e.g.~\cite{Cucchieri:1995pn}. Fixing to the $\max B$ gauge has been
discussed in section \ref{sec:landauB}. Propagators are determined
using a standard method, see, e.g., \cite{Cucchieri:2006tf}. For the
determination of the ghost propagator we also refer to Appendix
\ref{app:source}. A cylinder cut has been imposed to select momenta
\cite{Leinweber:1998uu}.
 
\subsection{Two dimensions}
\label{subsec:results2d}

\begin{figure}[t]
\begin{center}
\includegraphics[clip,width=0.92\columnwidth]{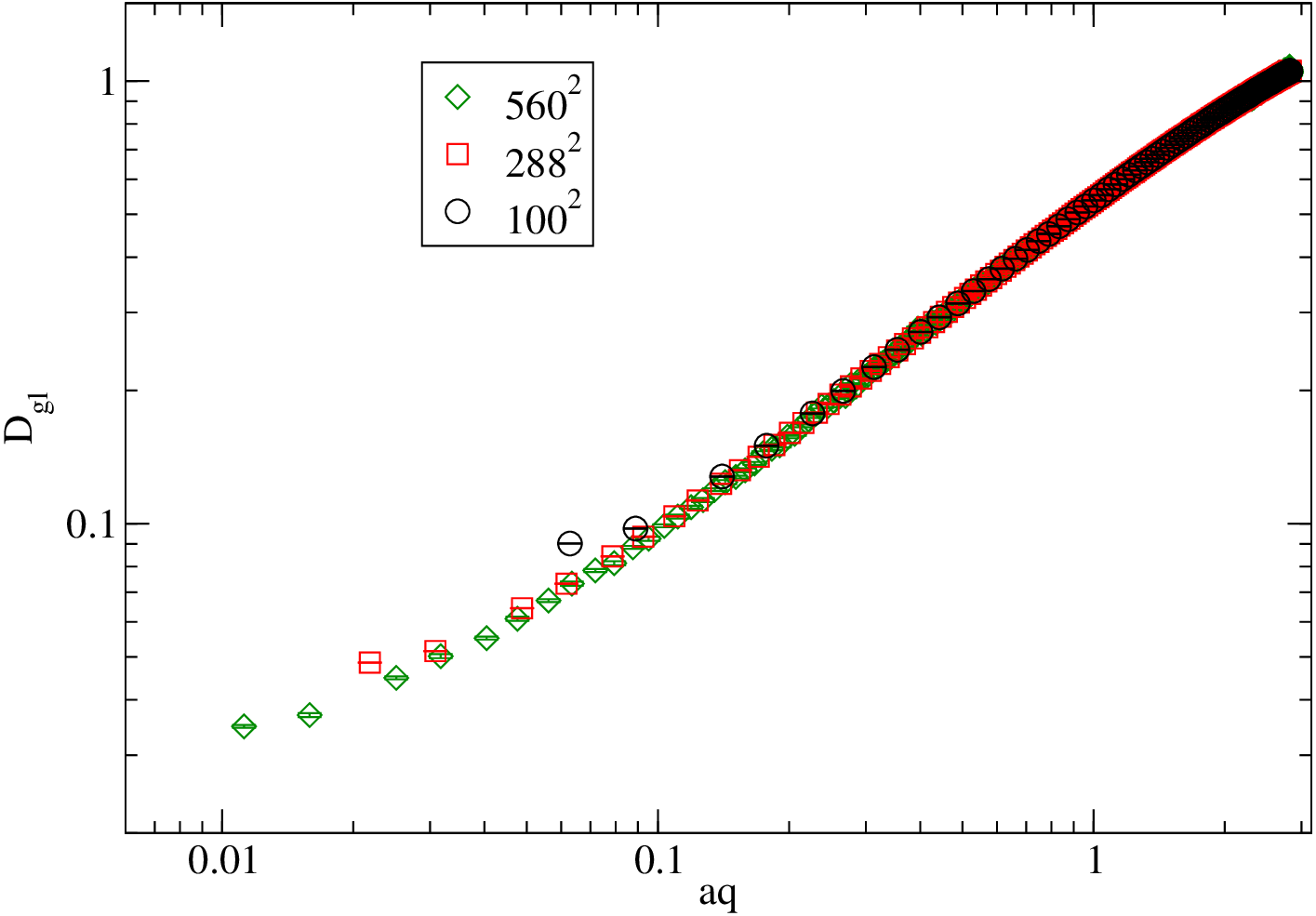}
\caption{\label{fig:glp_beta0_varls_2d_forpaper}Gluon propagator in
  $d=2$ for different lattice sizes.}
\end{center}
\end{figure}

\begin{figure}[t]
\begin{center}
\includegraphics[clip,width=0.92\columnwidth]{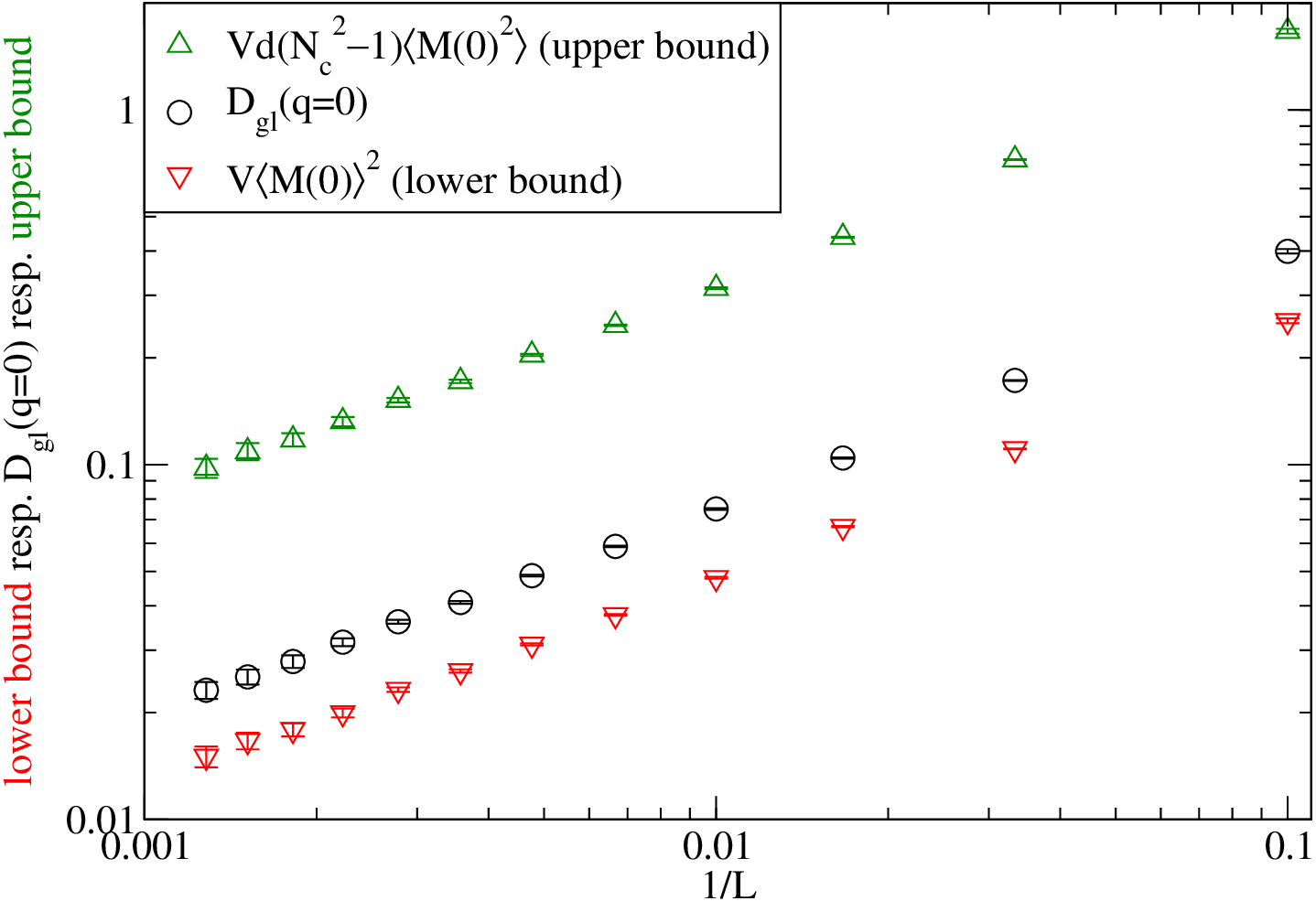} 
\caption{\label{fig:D0_beta0_2d}Finite-volume behavior of the gluon
  propagator at zero momentum in $d=2$. Also shown are the
  corresponding bounds from \cite{Cucchieri:2007rg}.}
\end{center}
\end{figure}

The gluon propagator for various lattice sizes in $d=2$ is shown in
\Fig{fig:glp_beta0_varls_2d_forpaper}. Its value at vanishing momentum
continues to decrease with the inverse lattice extension
(\Fig{fig:D0_beta0_2d}), and therefore only a lower bound on a
possible mass at zero momentum can be given. Power-law fits to the
gluon propagator, according to \Eq{eq:infrared-gl}, yield a value of
$\kappa_Z\approx 0.19$ at intermediate lattice momenta $aq$ around 1,
i.e.~close to the scaling prediction of $\kappa_Z = 0.2$, with a
tendency to decrease towards a value of $0.16$ at large $aq$. At this
point, this would all still be quite consistent with the scaling
solution \cite{Zwanziger:2001kw}, which has also been concluded from
previous lattice studies at finite coupling \cite{Maas:2007uv}.

Our results for the ghost propagator, however, show much more
significant deviations from this behavior. In particular, it appears
that they cannot be reconciled with the infrared scaling hypothesis,
as widely expected to hold at least for the two-dimensional theory
\cite{Dudal:2008xd,Cucchieri:2007rg,Cucchieri:2008fc,%
  Maas:2007uv,Pawlowski:2009iv,Zwanziger:2001kw,Lerche:2002ep}.  In order to draw this
conclusion, it is first of all crucial to have a sound statistical
basis. In that respect it made a dramatic difference that we have
employed a plane-wave method in obtaining the ghost propagator from
the inversion of the Faddeev--Popov operator, instead of the
point-source method used in \cite{Cucchieri:2009zt}. The two methods
are compared and discussed in more detail in Appendix
\ref{app:source}.

\begin{figure}[t]
  \includegraphics[clip,width=0.92\columnwidth]{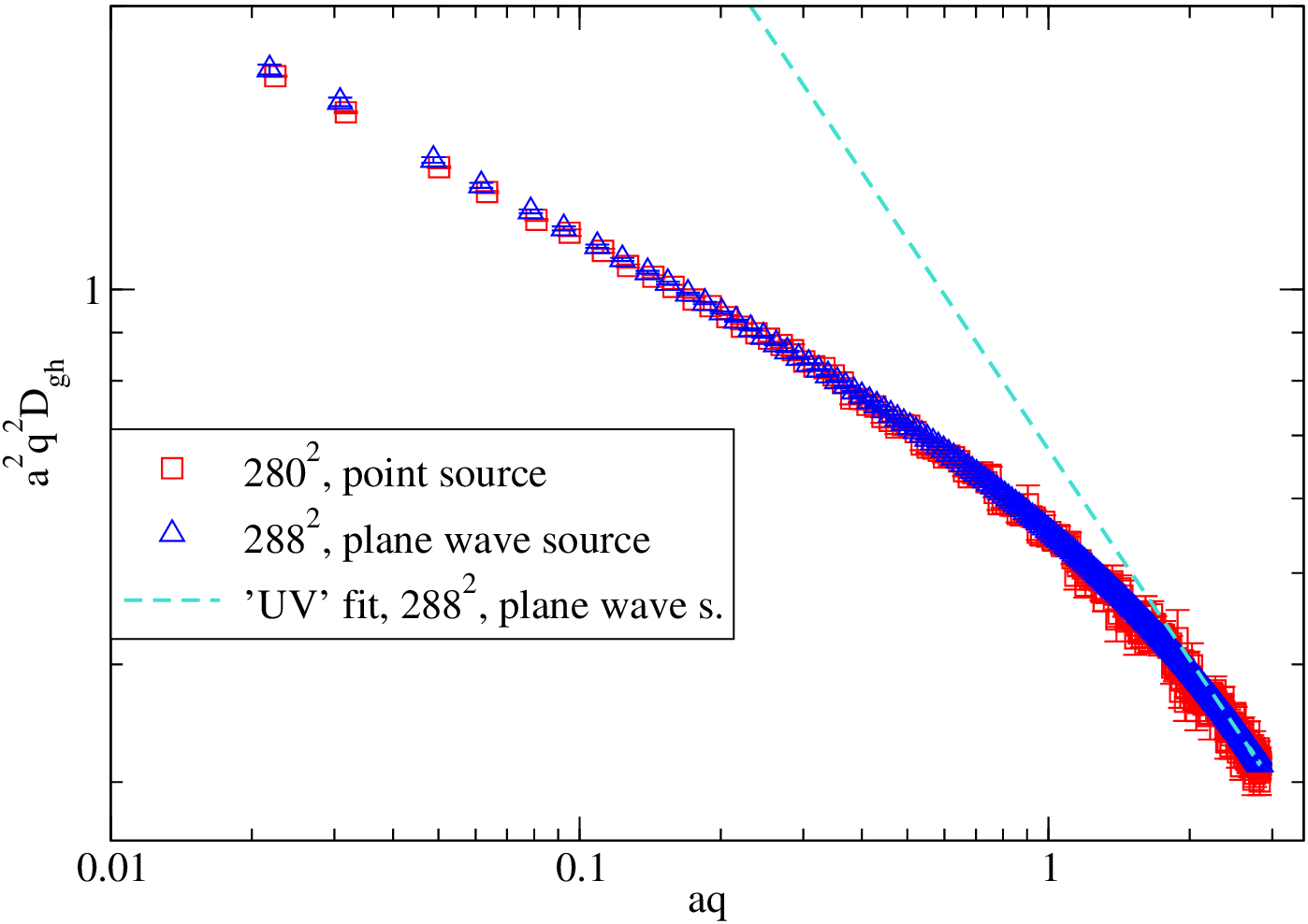} 
  \caption{\label{fig:ghdf_beta0_varls_2d_forpaper}Ghost dressing
    function in $d=2$.}
\end{figure}

\begin{figure}[t]
\begin{center}
\vspace{0.5cm}
  \includegraphics[clip,width=0.92\columnwidth]{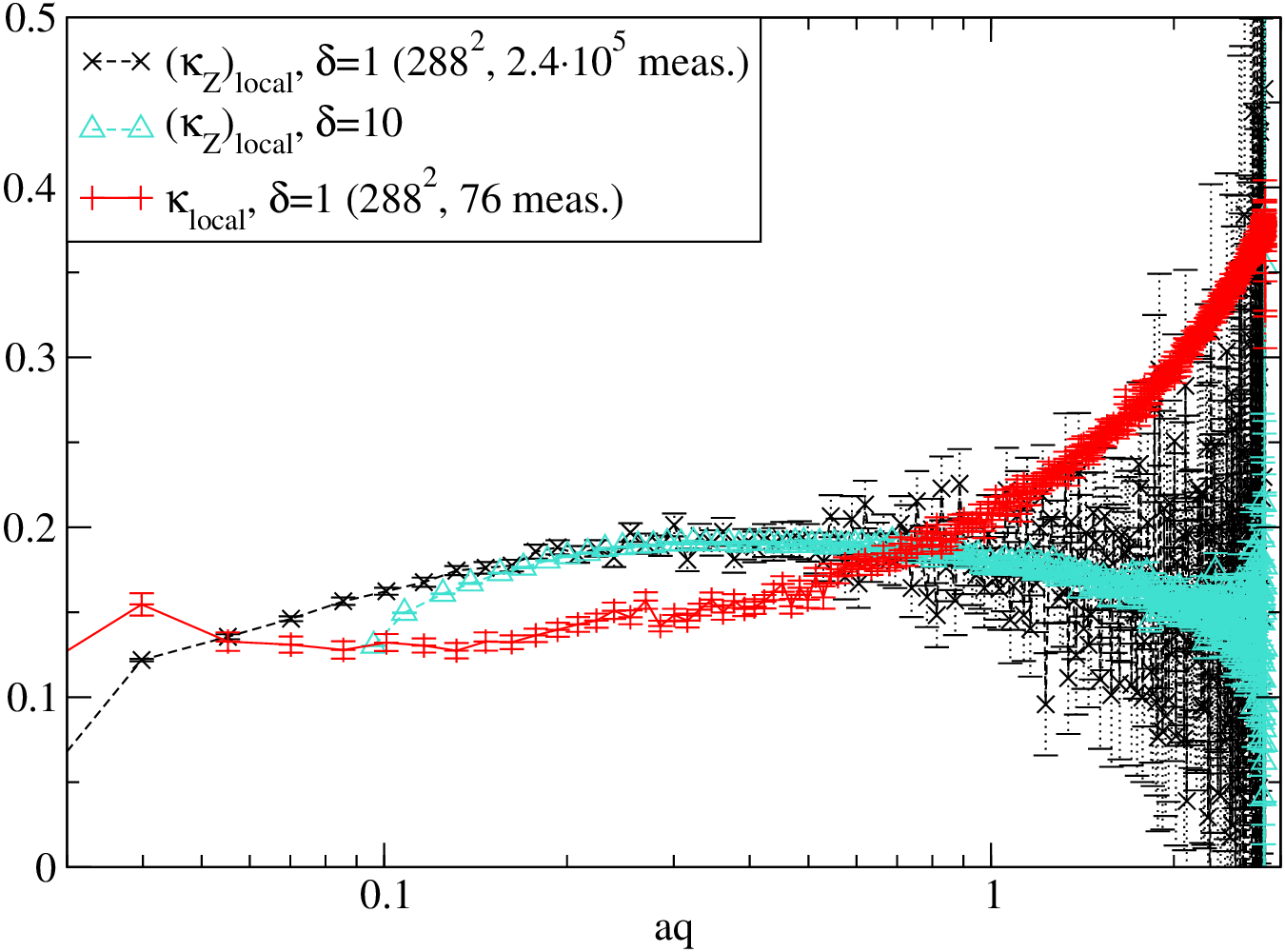} 
  \caption{\label{fig:averagekappa_glpetghp_ls288_2d_varwl}`Local
    $\kappa$', i.e.\ extracted from $q_i$ and $q_{i+\delta}$ from the 2D
    gluon ($\kappa_Z$) and ghost ($\kappa$) data, via Eqs.~\eqref{eq:Kzlocal2D}
    and \eqref{eq:Klocal2D}.}
\end{center}
\end{figure}

Fitting the ghost propagator, shown in
\Fig{fig:ghdf_beta0_varls_2d_forpaper}, with a naive power-law ansatz
at intermediate and large momenta, we extract a value of around
$\kappa=0.37$.  This is significantly larger than the expected value
of $0.2$ \cite{Zwanziger:2001kw,Maas:2007uv} and than the
corresponding scaling exponent from the two-dimensional gluon
propagator in this range.  Moreover, the $\kappa$ from the ghost
propagator continues to increase as larger and larger lattice momenta
are being used for the fit. A good way to illustrate this, and thus
the absence of a stable scaling region for the ghost propagator in $d=2$
dimensions, is to extract the value of $\kappa$ locally, from just two
momenta (i.e., momentum norms) via
\begin{subequations}
\begin{align}
  \label{eq:Kzlocal2D}
  (\kappa_Z)_{\mathrm{local}}& =\frac{1}{4}
  \left[\frac{\log D_\text{gl}(q_{i+\delta})-\log
      D_\text{gl}(q_i)}{\log q_{i+\delta}-\log q_i} +(d-2)\right]\\
\label{eq:Klocal2D}
  (\kappa)_{\mathrm{local}}&=\frac{1}{2}\left[\frac{\log D_\text{gh}(q_i)
      -\log D_\text{gh}(q_{i+\delta})}{ \log q_{i+\delta} -\log q_i}-2\right]\,;
\end{align}
\end{subequations}
see \Fig{fig:averagekappa_glpetghp_ls288_2d_varwl}. The statistical errors
have been estimated using a bootstrap analysis.\footnote{A similar
  analysis was performed in three dimensions (see also our next
  subsection) already in \cite{Cucchieri:2009zt} with the important
  difference, however, that the Faddeev--Popov operator was there
  inverted on a point source.} If anywhere, the locally defined
exponent levels around relatively small lattice momenta of $aq$
around 0.1 at a value which is several standard deviations below the
gluon exponent there, and it continues to grow from there on towards
larger momenta.  Our results are clearly at odds with results at
finite coupling on moderately sized lattices \cite{Maas:2007uv}
supposedly showing a scaling behavior in agreement with predictions
\cite{Zwanziger:2001kw,Lerche:2002ep}.

\subsection{Three dimensions}
\label{subsec:results3d}

\begin{figure}[t]
\begin{center}
  \includegraphics[clip,width=0.92\columnwidth]{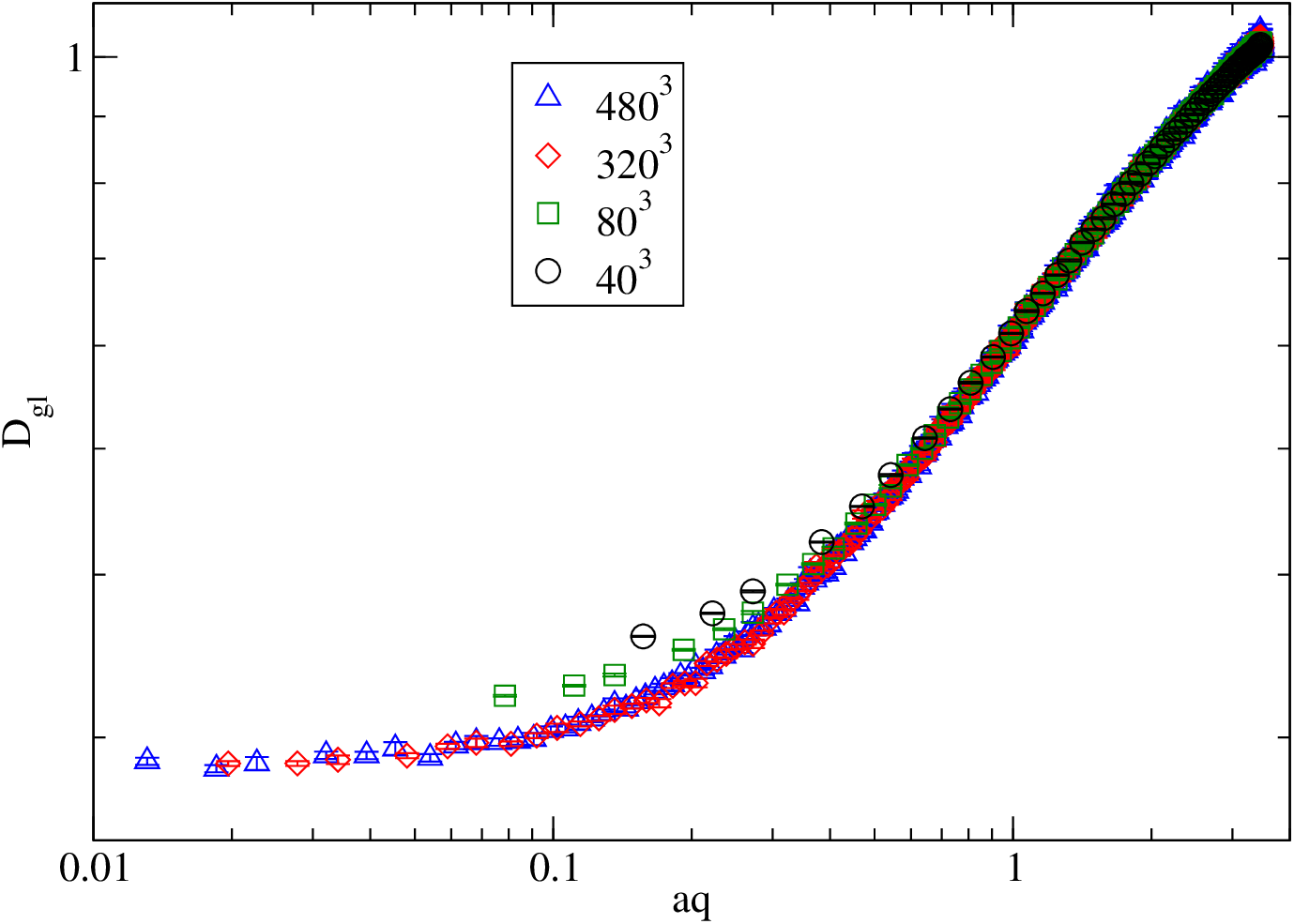}
  \caption{\label{fig:glp_beta0_bc0_varls_loglog_forpaper}Gluon
    propagator in $d=3$ for various lattice sizes.}
\end{center}
\end{figure}

\begin{figure}[t]
\begin{center}
  \includegraphics[clip,width=0.92\columnwidth]{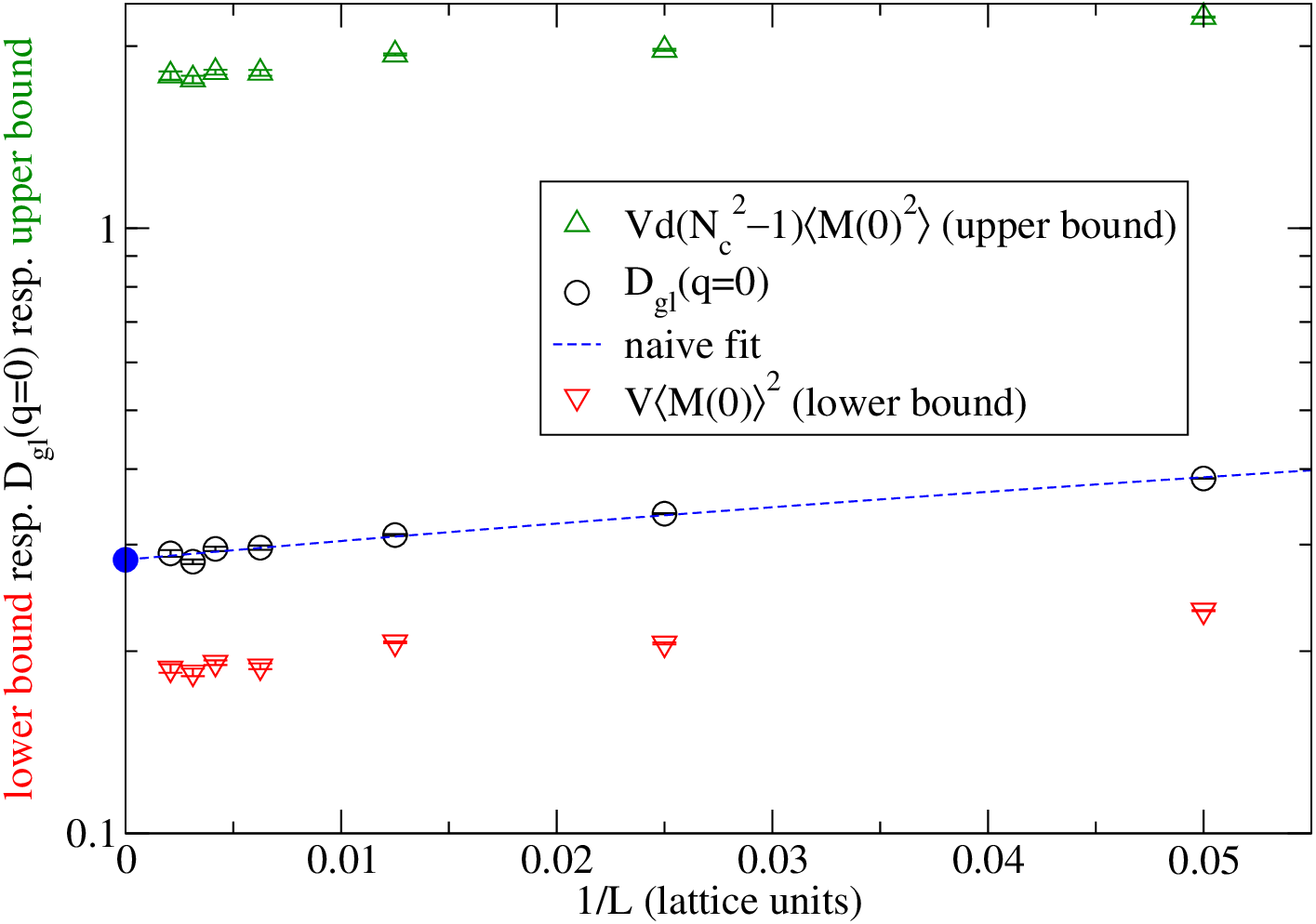}
  \caption{\label{fig:D0_forposter}Naive extrapolation of the gluon
    propagator in $d=3$ at zero momentum to infinite lattice size. The
    decoupling branch persists in this limit. The bounds from
    \cite{Cucchieri:2007rg} are also indicated.}
\end{center}
\end{figure}

The gluon propagator in $d=3$ is again relatively well-behaved. It
shows a decoupling branch at small lattice momenta $aq$ and a behavior
resembling a scaling branch at large $aq$, see
\Fig{fig:glp_beta0_bc0_varls_loglog_forpaper}. While the decoupling
branch still shows quite significant finite-volume effects, these are
clearly not strong enough to permit a vanishing gluon propagator at
zero momentum in the $L\to\infty$ limit, see \Fig{fig:D0_forposter}.

\begin{figure}
\begin{center}
\includegraphics[clip,width=0.92\columnwidth]{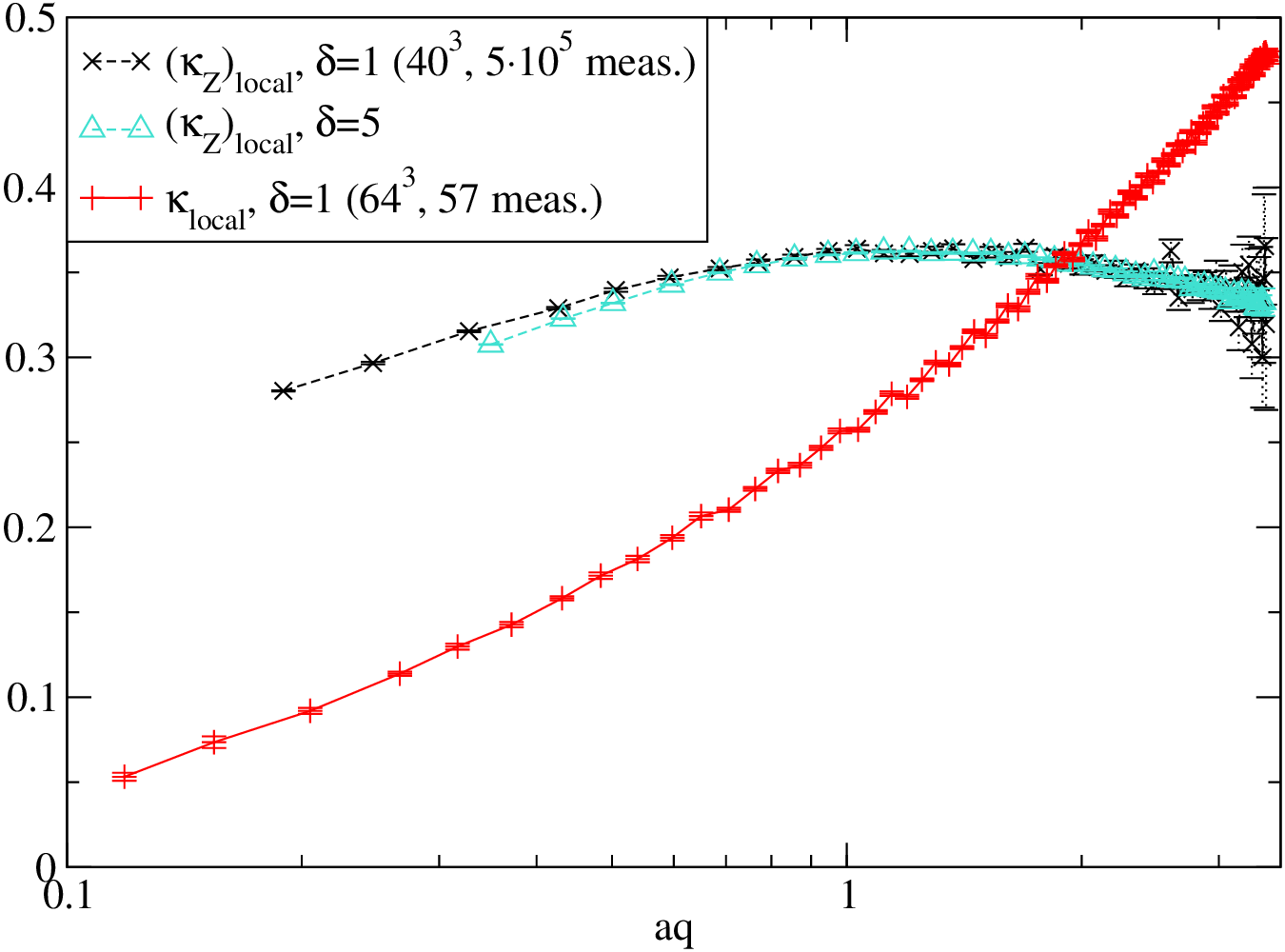}
\caption{\label{fig:averagekappa_ghdf_plane_beta0_ls64_3d_fp} `Local
    $\kappa$', as in \Fig{fig:averagekappa_glpetghp_ls288_2d_varwl}
    but for three dimensions.}
\end{center}
\end{figure}

As in two and four dimensions, the branch at large $aq$ is consistent
with a power law with an exponent $\kappa_Z$ close to the value
predicted by functional continuum methods, i.e., $\kappa_Z\approx
0.35$ versus the predicted $0.3976 $ \cite{Zwanziger:2001kw}, and in
agreement with \cite{Cucchieri:2009zt}. At large lattice momenta, the
power-law exponent slightly decreases again, which is also evident
from its local definition, see
\Fig{fig:averagekappa_ghdf_plane_beta0_ls64_3d_fp}.

\begin{figure}[t]
\begin{center}
 \includegraphics[clip,width=0.92\columnwidth]{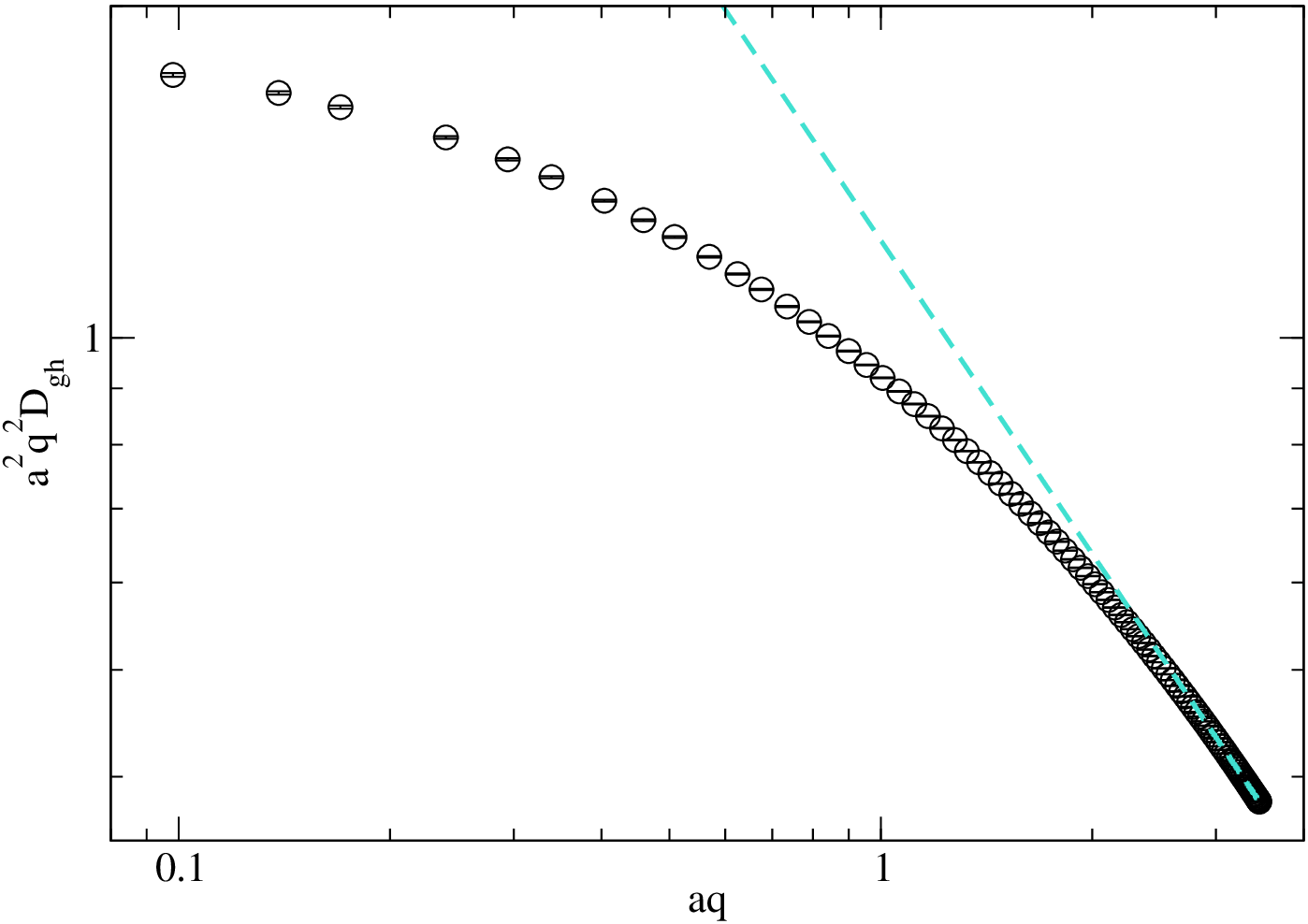}
 \caption{\label{fig:ghdf_beta0_plane_64_withfit}Ghost dressing
   function for a $64^3$ lattice. The dashed line is a power-law fit
   to large momenta.}
\end{center}
\end{figure}

At the same time, \Fig{fig:averagekappa_ghdf_plane_beta0_ls64_3d_fp}
illustrates that things become murky again when analyzing the ghost
propagator for potential power-law windows. Qualitatively, the
situation is similar to that in $d=2$ dimensions, the locally defined
exponent continues to increase with increasing momenta and the ghost
propagator cannot reasonably be interpreted as obeying a scaling
behavior in any momentum range. The result for the ghost propagator
itself is given in \Fig{fig:ghdf_beta0_plane_64_withfit} which
includes a fit over a maximal data set at large lattice momenta
(extending up to the largest ones) such that $\chi^2/\ndf<1$. The
general conclusions are much the same as in the $d=2$ case above.

\subsection{The $\max B$  gauge}\label{res:maxb}

\begin{figure*}
 \includegraphics[clip,width=0.92\columnwidth]{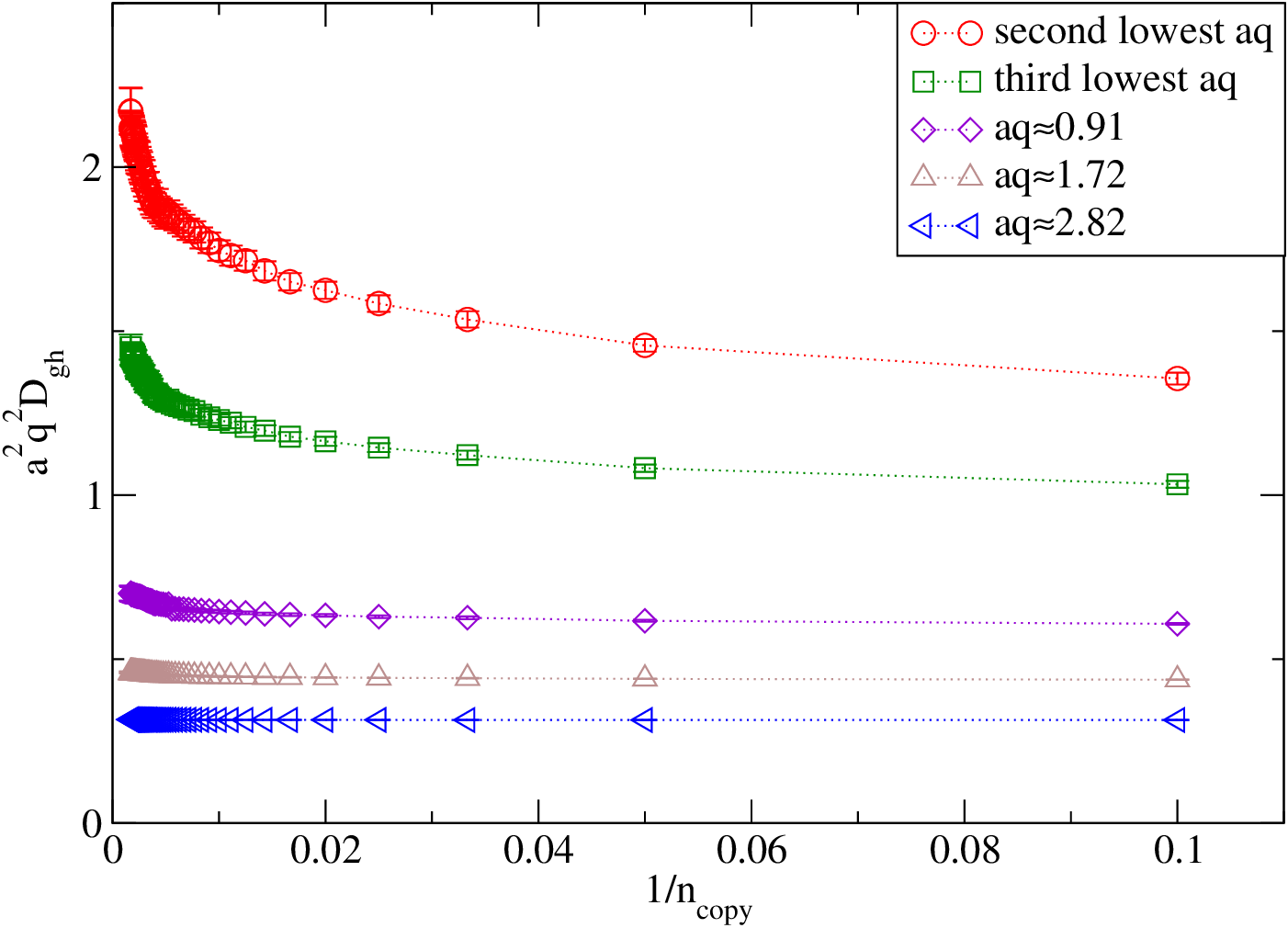}\qquad\qquad
 \includegraphics[clip,width=0.92\columnwidth]{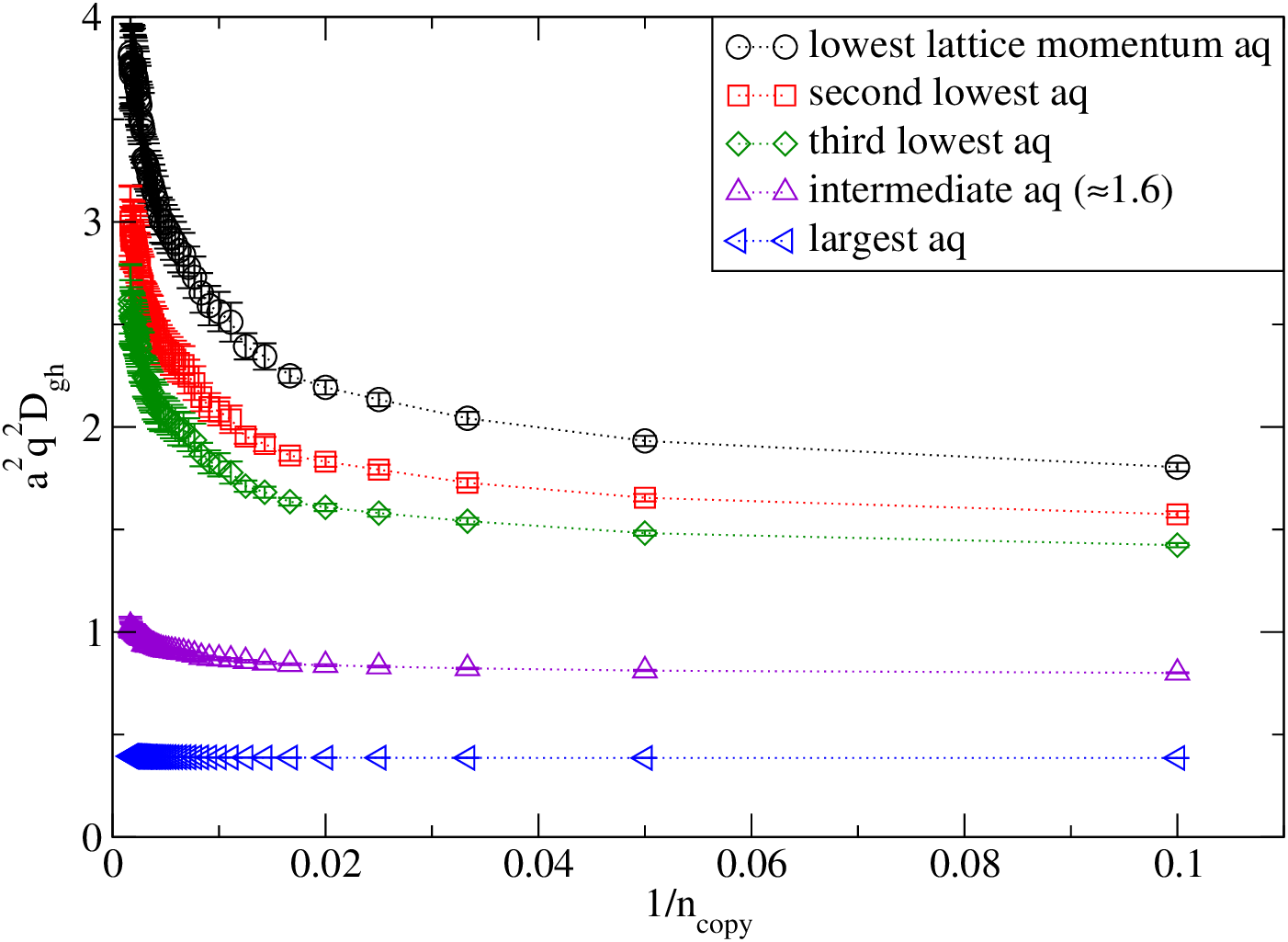} 
\caption{\label{fig:ghdf_DIV_vs_axgf}Ghost dressing function vs.\
  inverse number of Gribov copies for the $\max B$ gauge. Left: for
  $d=2$ dimensions and a $48^2$ lattice. Right: for $d=3$ dimensions
  and a $20^3$ lattice. Lines are drawn to guide the eye.} 
\end{figure*}

The main aspect of the results that prevents a straightforward
interpretation is the behavior of the ghost propagator which is most
sensitive to the Gribov ambiguity. It therefore seems to be a natural
question to ask what the influence is of this residual gauge freedom
of the Landau gauge which can be used to change the ghost propagator
as explained in Sec.~\ref{sec:landauB}. Here we deliberately choose
the $\max B$ Landau gauge of Sec.~\ref{sec:landauB} in order to
demonstrate the maximal effect. These are the first calculations at
$\beta=0$ in this gauge. In the strong-coupling limit one generates
random gauge orbits for which the typical number of Gribov copies is
extremely large. Moreover, this number is expected to grow
exponentially with the number of lattice sites $N^d$. It will
therefore be practically impossible to find even a significant
fraction of all copies on large lattices. We therefore restrict the
analysis to relatively modest lattice sizes such as $48^2$ and
$20^3$. It would have been possible of course that the results
saturated rather quickly with the number of Gribov copies taken into
account, as it appeared to be the case in calculations at $\beta>0$
\cite{Sternbeck:2005tk,Silva:2004bv}. This does not happen here,
however, at least not for the up to 600 copies per orbit analyzed on
these lattices. As one can see below in \Fig{fig:ghdf_DIV_vs_axgf}, in
particular the low-momentum values of the ghost propagator continue to
grow as more and more Gribov copies are included in the $\max B$ gauge.

\begin{figure}[t]
\begin{center}
\includegraphics[clip,width=0.92\columnwidth]{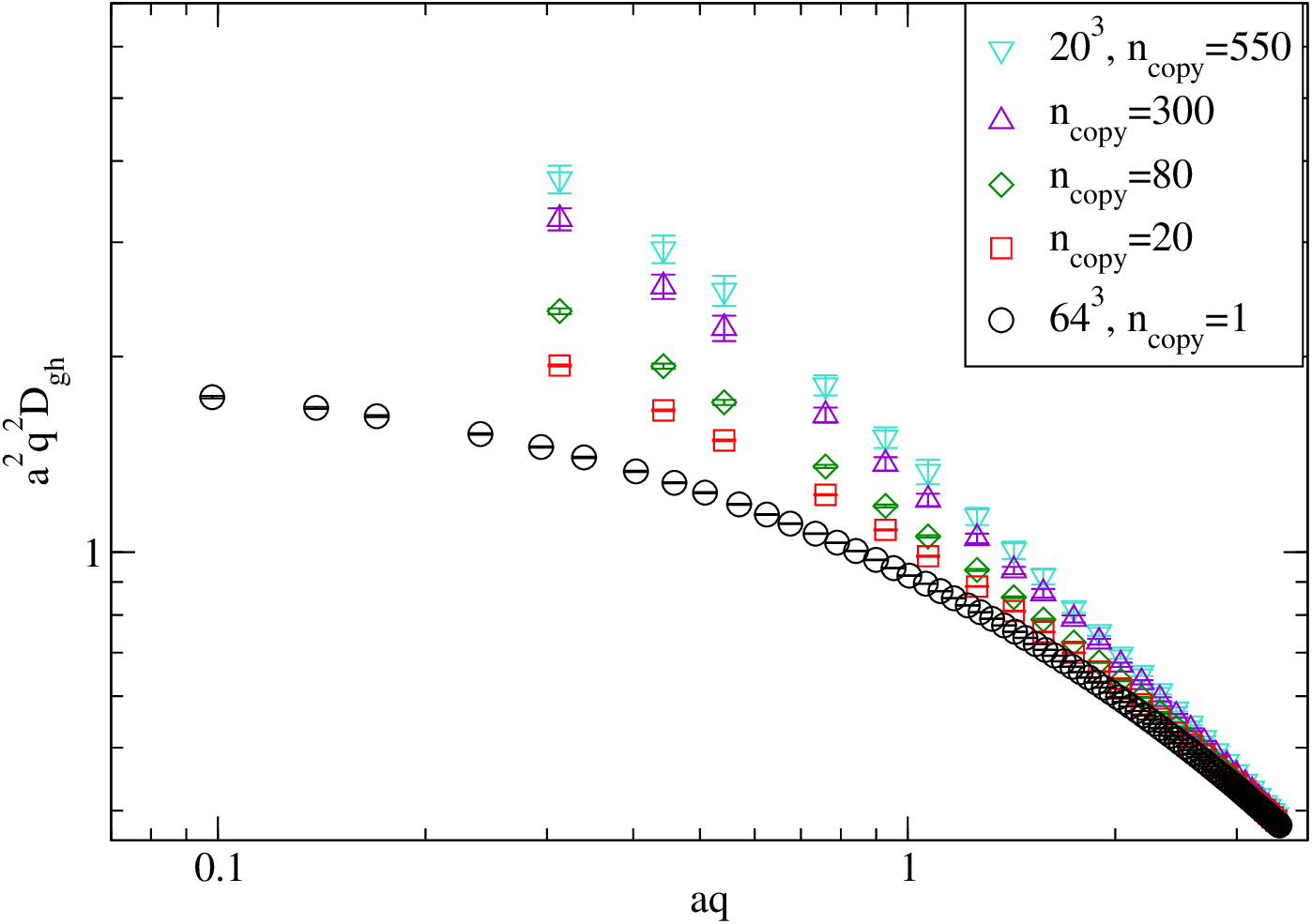}
\caption{\label{fig:ghdf_beta0_plane_noaxgfvsaxgf20_varl}Ghost
  dressing function in the $\max B$ gauge for different numbers of
  Gribov copies taken into account.}
\end{center}
\end{figure}

\begin{figure}[t]
\begin{center}
\includegraphics[clip,width=0.92\columnwidth]{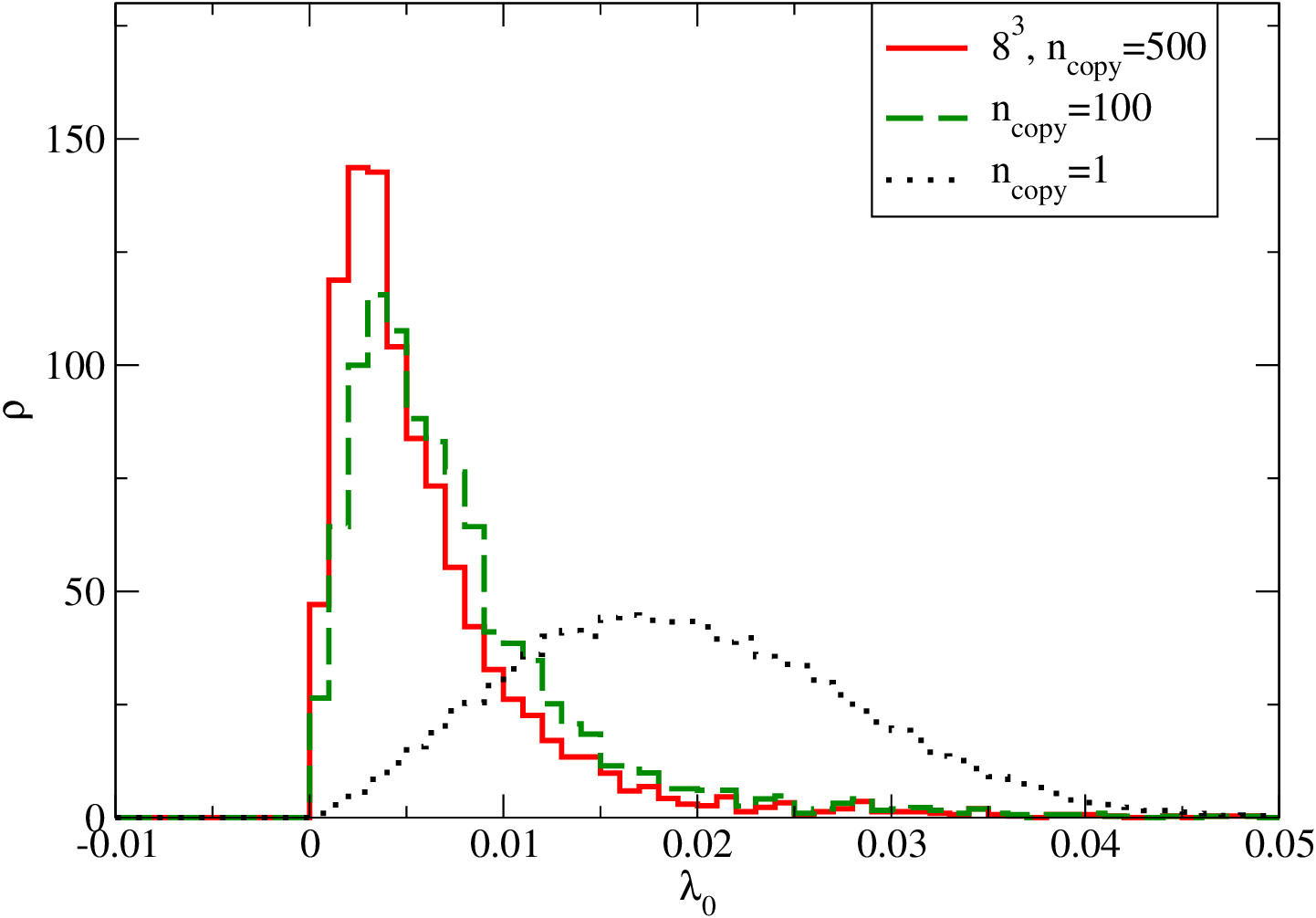} 
\caption{\label{fig:FPO_merged_ls8_varnaxgf}Effect of the $\max
  B$-gauge on the lowest nontrivial (i.e., non-vanishing) FPO
  eigenvalue as a function of included Gribov copies. Entries in the
  histogram correspond to the lowest eigenvalue on different
  configurations for the Gribov copy with the most divergent ghost
  propagator.}
\end{center}
\end{figure}

\begin{figure}
\begin{center}
\includegraphics[clip,width=0.92\columnwidth]{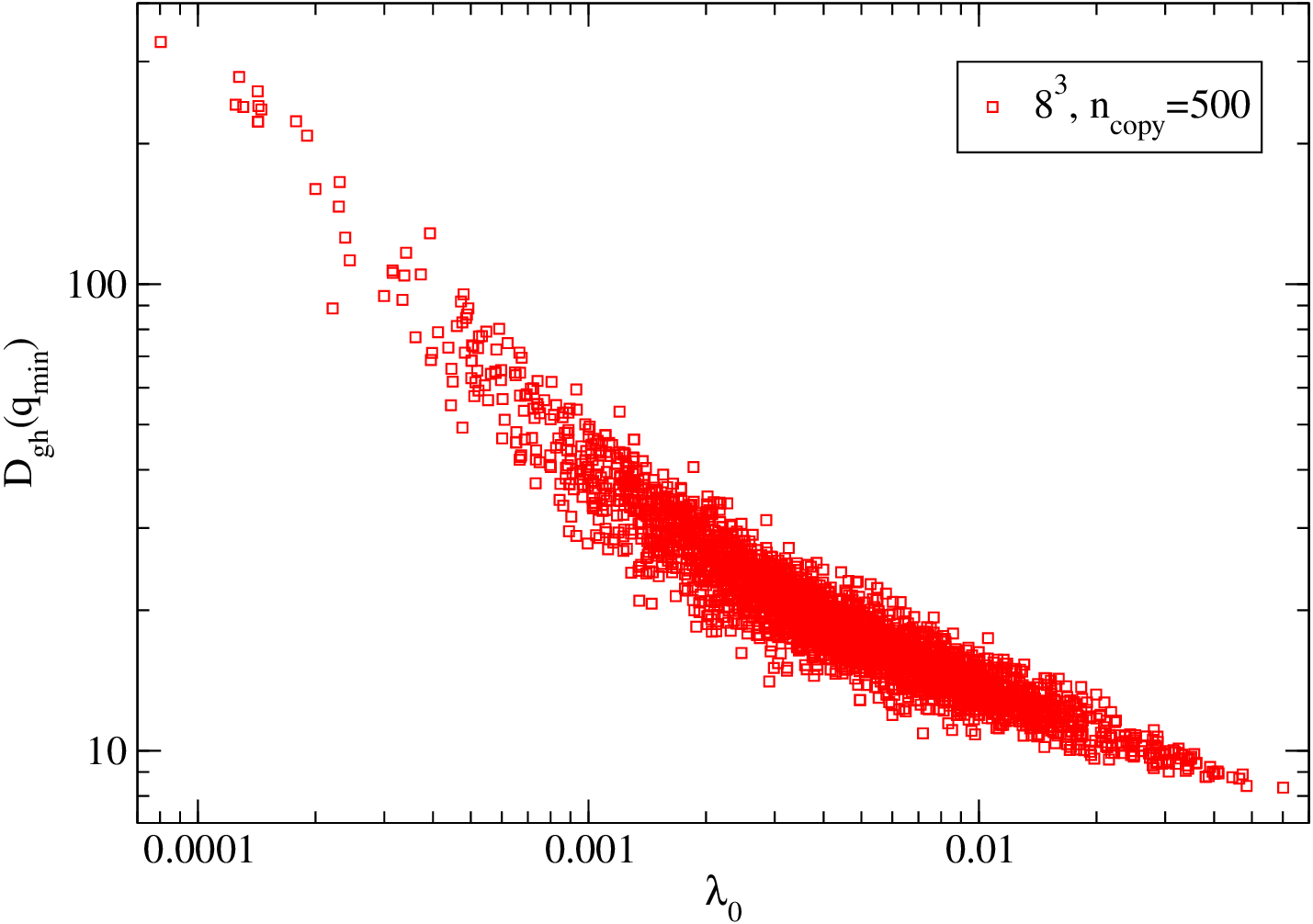} 
\caption{\label{fig:lambda0-ghp-8-3d-n500}Strong correlation between
  lowest FPO eigenvalue and ghost propagator at lowest non-vanishing
  momentum (in the $\max B$ gauge).}
\end{center}
\end{figure}

Our results for the ghost propagator in $d = 3 $ dimensions are shown
in \Fig{fig:ghdf_beta0_plane_noaxgfvsaxgf20_varl}.  We observe a
strong increase in the ghost dressing function with the number of
Gribov copies $n_\text{copy}$ included per orbit in the search for the
$\max B$ gauge. The observed enhancement in the ghost dressing
function at low momenta comes along with a significant shift in the
spectrum of the lattice Faddeev--Popov operator (FPO), as one would
expect. This is illustrated on an even smaller lattice in
\Fig{fig:FPO_merged_ls8_varnaxgf}. The strong correlation between the
lowest Faddeev--Popov eigenvalue $\lambda_0$, omitting the
three trivial zero eigenmodes, and the ghost propagator at lowest
non-vanishing momentum is seen in \Fig{fig:lambda0-ghp-8-3d-n500} (see
also \cite{Sternbeck:2005vs}).

\begin{figure}
\begin{center}
\includegraphics[clip,width=0.92\columnwidth]{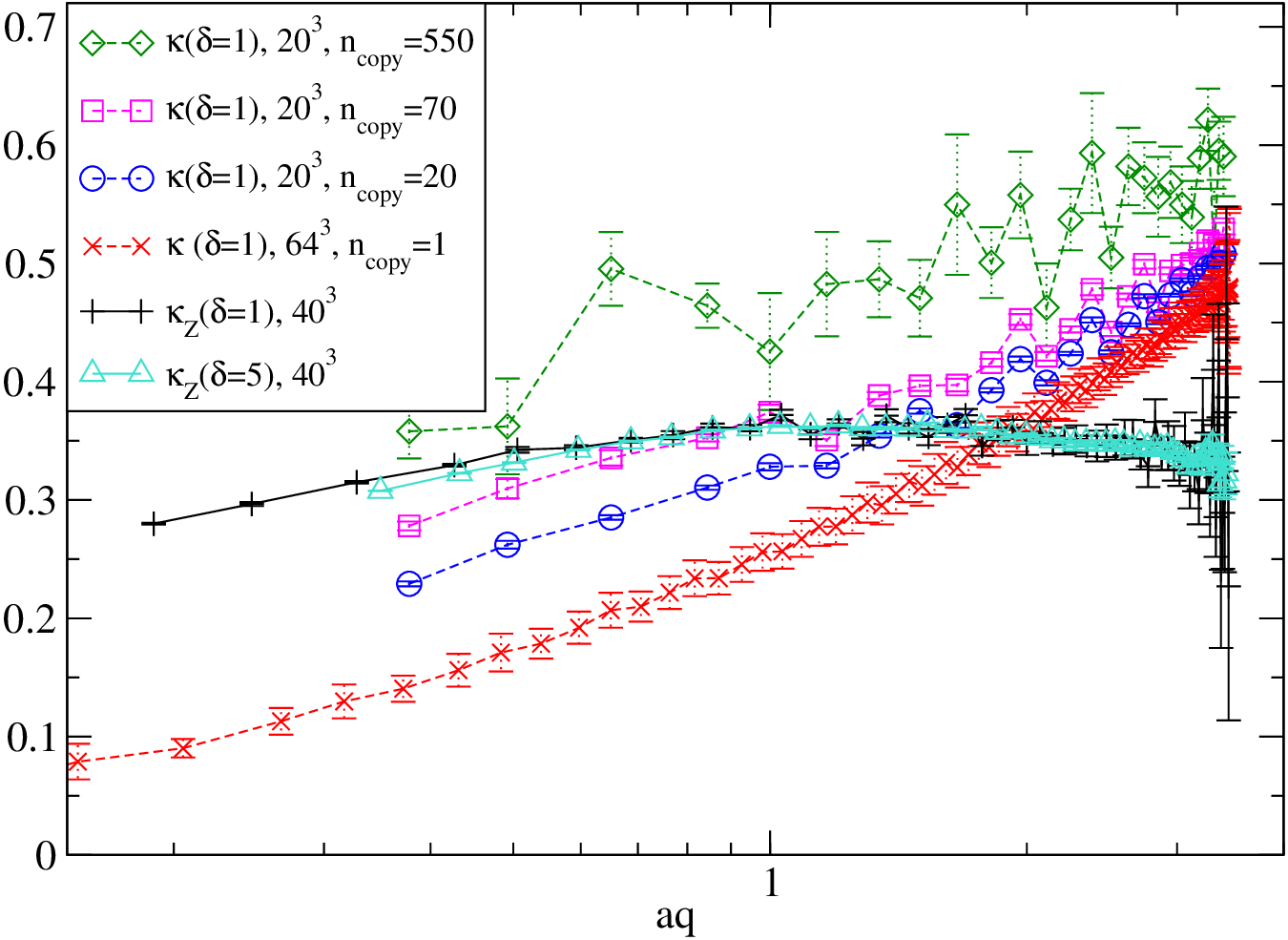}
\caption{\label{fig:averagekappa_glpetghp_axgf20vsnoaxgf_3d}Local
  $\kappa_Z$ resp. $\kappa$ from the $\max B$ gauge.}
\end{center}
\end{figure}

It seems surprising that the ghost propagator in the $\max B$ gauge
appears to show no sign of saturation with the number of Gribov copies
taken into account, even on these small lattices and in both, two and
three dimensions, as illustrated in
Fig.~\ref{fig:ghdf_DIV_vs_axgf}. It appears not
likely that the ghost propagator should increase without bound as the
number of copies is further increased in a finite volume, but we are
obviously still far from having sampled sufficiently many to conclude
that we could reach a stable limit. We conclude that Gribov copies do
have an enormous effect on the low-momentum behavior of the ghost
propagator.  This is also reflected in the corresponding $\kappa$'s as
shown in Figure \ref{fig:averagekappa_glpetghp_axgf20vsnoaxgf_3d}.

It should be noted that by choosing instead of the maximum value of
the ghost propagator a finite value for $B$, it is possible to
generate any ghost propagator in between the one of minimal Landau
gauge and the one in the $\max B$ gauge. In fact, it is also possible
to select a ghost propagator which is smaller than the one in minimal
Landau gauge \cite{Maas:2009se}.

However, our primary interest in employing the $\max B$ gauge is to
investigate a potential fixed-point behavior in the running coupling,
an issue to which we now turn.
\subsection{Running coupling}\label{res:maxbrc}

\begin{figure}
\begin{center}
\includegraphics[clip,width=0.92\columnwidth]{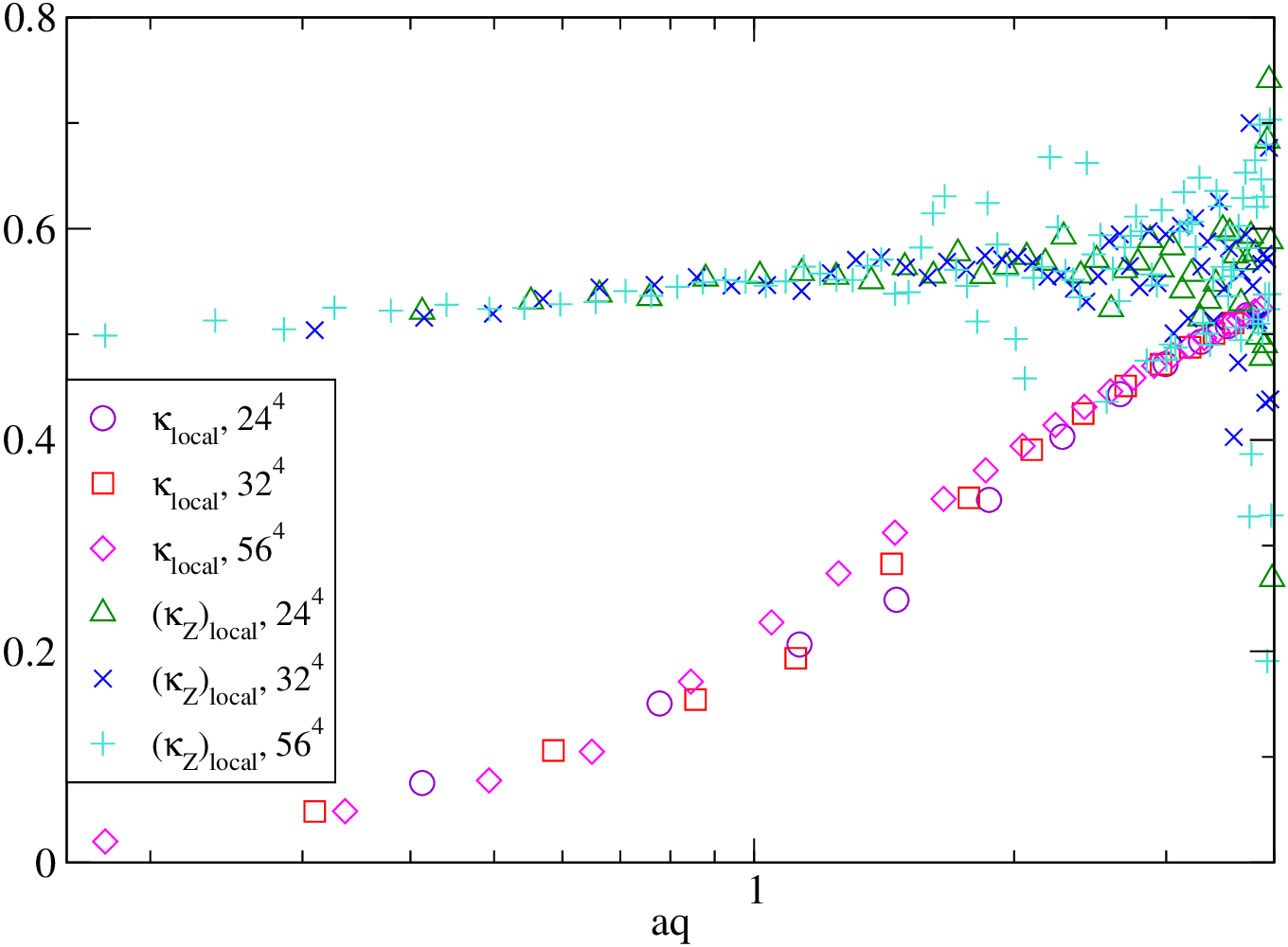}
\caption{\label{fig:kappa_kappaZ_eff_qq} `Local
    $\kappa$', as in \Fig{fig:averagekappa_glpetghp_ls288_2d_varwl}
    but for $d=4$ dimensions; see \cite{Sternbeck:2008mv} for the
    corresponding propagator results. For the ghost propagator, not
    all momenta eligible under the cylinder cut have been used
    here. The exponent $\kappa_Z$ from the gluon data is virtually
    constant except for the smallest lattice momenta. Error bars are
    relatively small and not shown.}
\end{center}
\end{figure}

Let us emphasize that an increase of $\kappa$ from the ghost
propagator towards large lattice momenta is actually also apparent
from the $d=4$ data presented already in
\cite{Sternbeck:2008wg,Sternbeck:2008mv,Sternbeck:2008na}, as shown
explicitly in \Fig{fig:kappa_kappaZ_eff_qq}. But it is important to
notice that in this case, unlike in $d=2$ and $d=3$,
\Fig{fig:ghdf_DIV_vs_axgf}, $\kappa$ from the
ghost data does not rise above $\kappa_Z$, i.e., the ghost exponent
extracted from the gluon data. This manifests itself in a different
behavior of the effective running coupling (\ref{eq:def-alphas}).  As
reported in \cite{Sternbeck:2008wg,Sternbeck:2008mv}, in $d=4$, $\alpha_s$ grows
monotonically as a function of $aq$ and approaches a value
$\alpha_c\approx 4$ from below, which is close to the value
$\alpha_c\approx 4.46$ predicted in \cite{Lerche:2002ep}.

\begin{figure}
\begin{center}

  \includegraphics[clip,width=0.92\columnwidth]{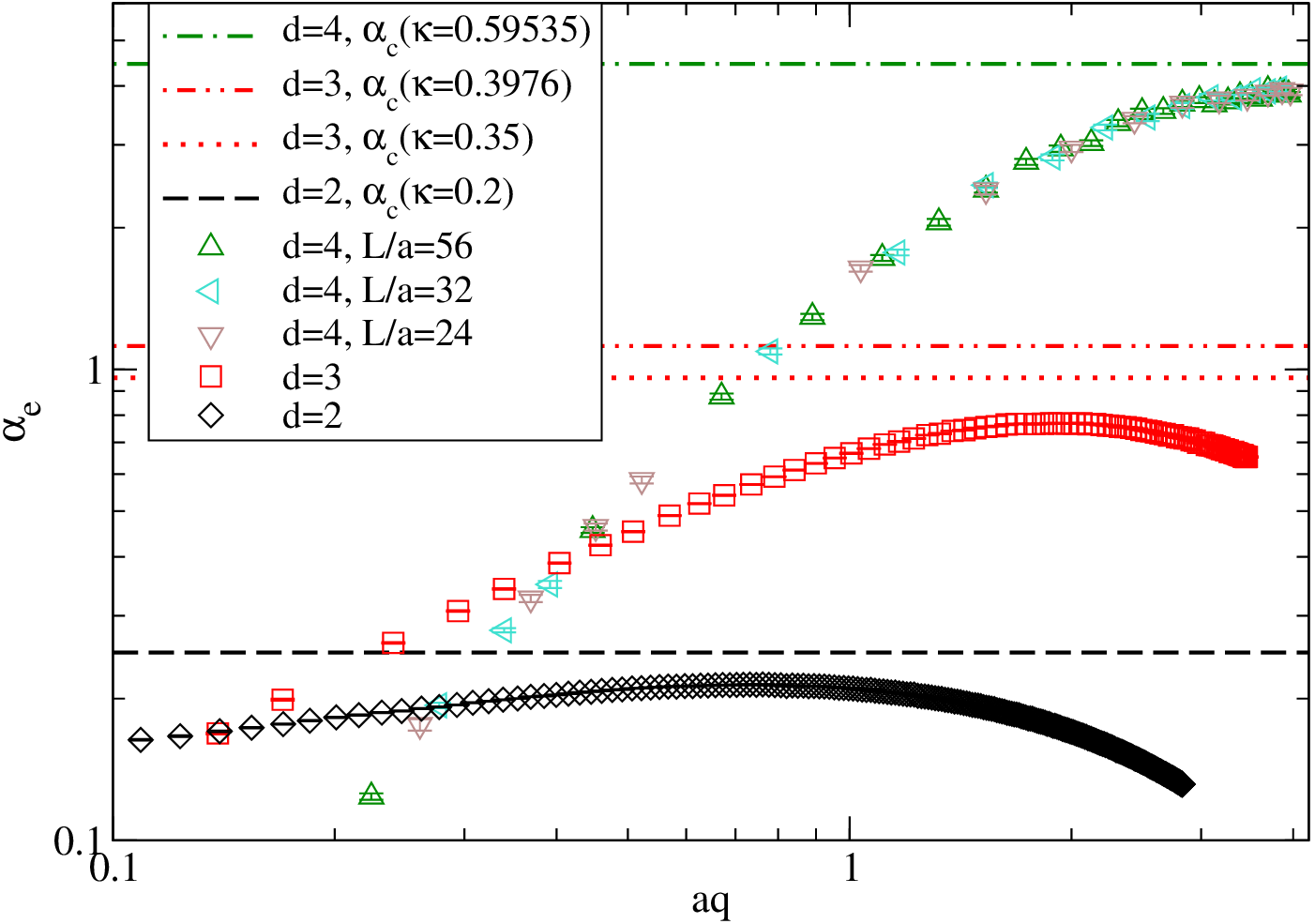}
  \caption{\label{fig:coup_forpaper}$\alpha_e$ in $d=2$, $L=288$, and
    $d=3$, $L=64$, as well as from various lattice sizes in $d=4$.}
\end{center}
\end{figure}

\begin{figure}
\begin{center}
  \includegraphics[clip,width=0.92\columnwidth]{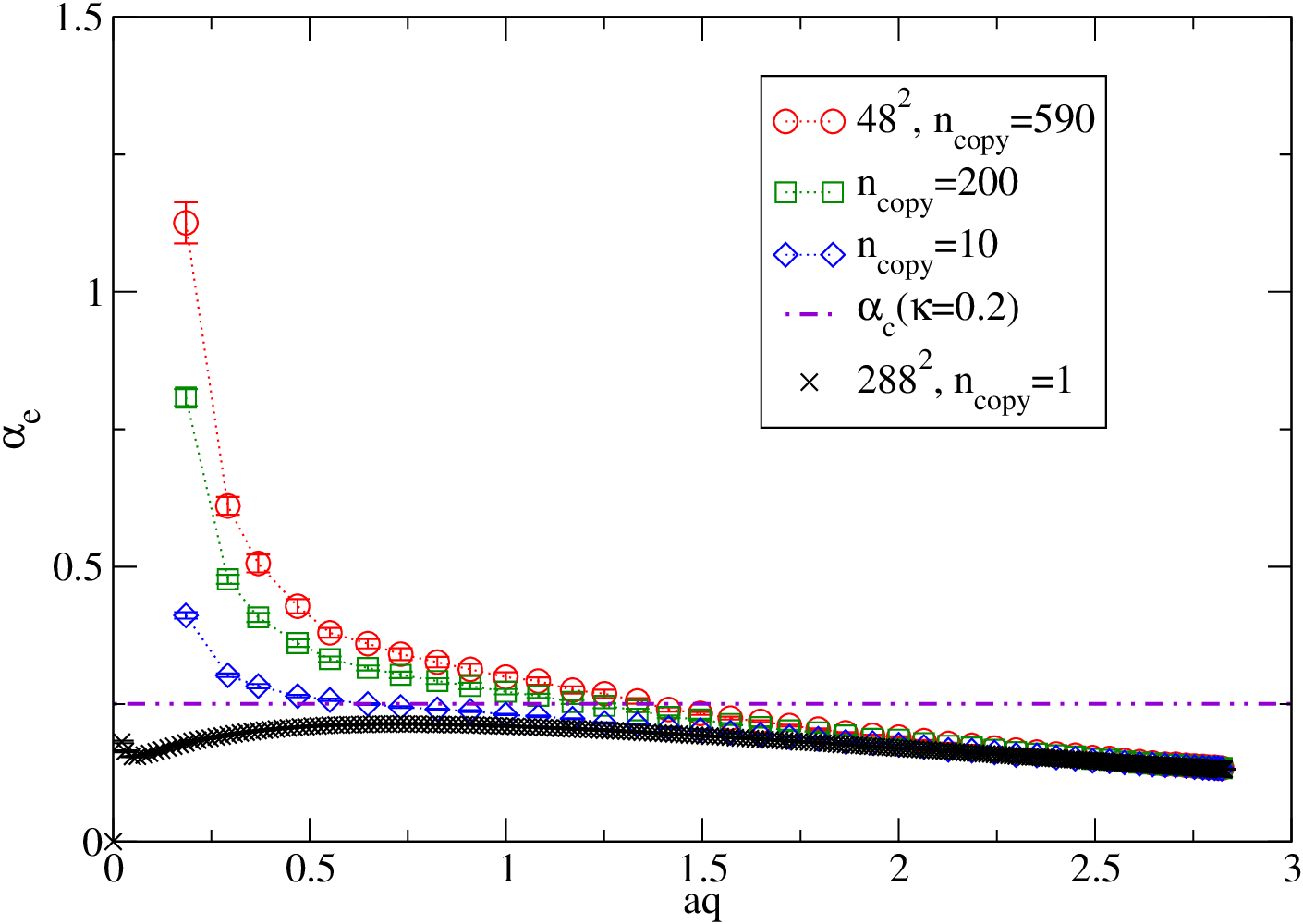}\\\vspace{0.2cm}\includegraphics[clip,width=0.92\columnwidth]{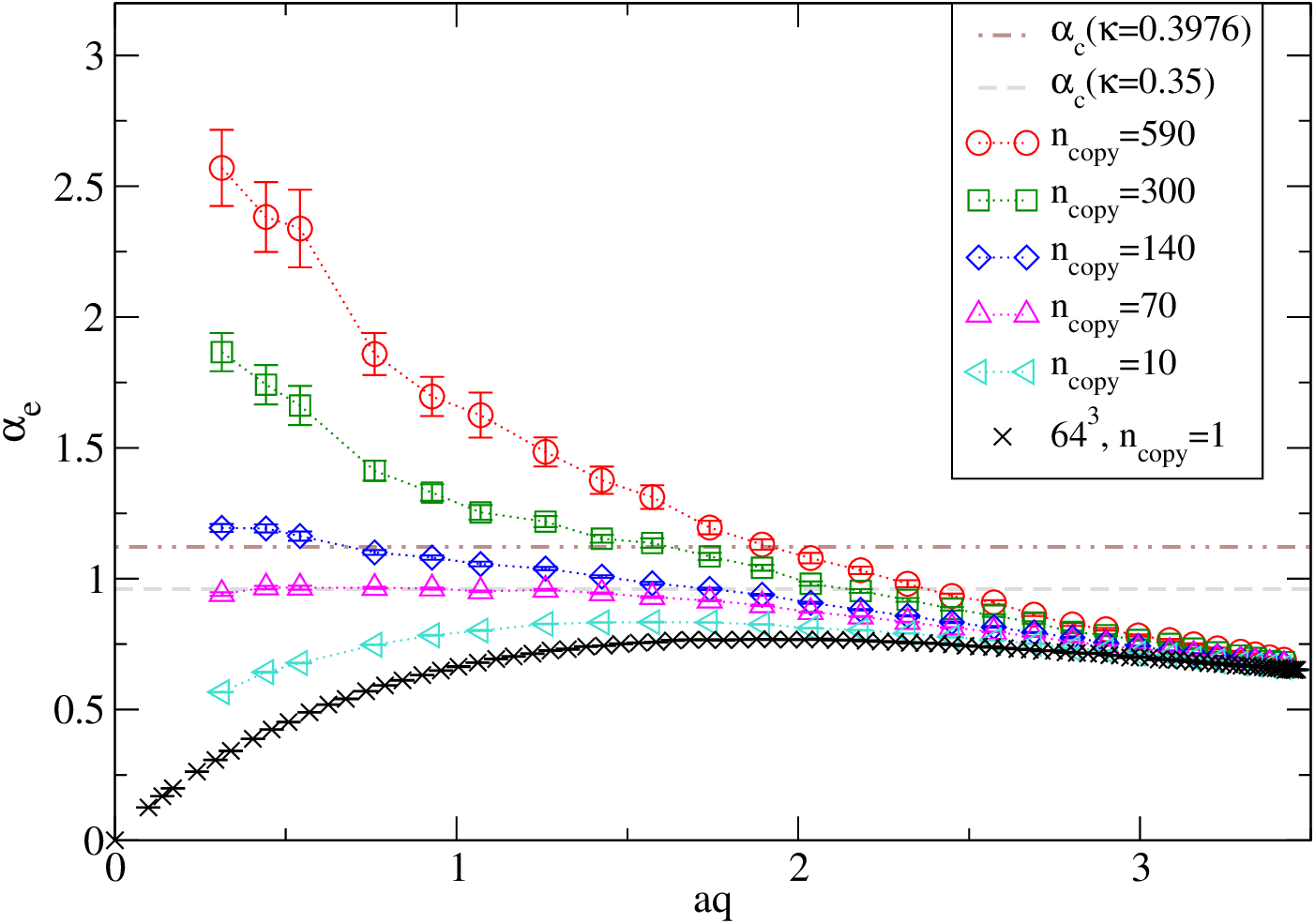}

\caption{\label{fig:coup_varnaxgf}$\max B$ gauge:
  Very strong change in the running coupling, especially at small lattice momenta, as the number of Gribov copies increases. Top: for $d=2$ dimensions and a $48^2$ lattice. Bottom: for $d=3$ dimensions and a $20^3$ lattice.}
\end{center}
\end{figure}

The behavior we observe here in minimal Landau gauge in $d=2$ and
$d=3$ differs surprisingly: The behavior at large lattice momenta is
notably different from the four-dimensional case, see
\Fig{fig:coup_forpaper}. This is obviously a joint consequence of the
behavior of $\kappa$ and $\kappa_Z$
(Figs.~\ref{fig:averagekappa_glpetghp_ls288_2d_varwl} respectively
\ref{fig:averagekappa_ghdf_plane_beta0_ls64_3d_fp}) and the fact that
from \Eq{eq:def-alphas}
\begin{equation}\label{alphas-kappa-kappaz}
\alpha_e\propto (q^2)^{2(\kappa_Z-\kappa)}.
\end{equation}
In $d=2$ and $3$, where the local $\kappa$ grows monotonically and
intersects $\kappa_Z$ at some $aq$, see
\Figs{fig:averagekappa_glpetghp_ls288_2d_varwl} and
\ref{fig:averagekappa_ghdf_plane_beta0_ls64_3d_fp}, $\alpha_e$ reaches
a maximum near the respective lattice momentum, as implied by
\Eq{alphas-kappa-kappaz}. Comparing the results to the predictions
from \cite{Lerche:2002ep}, we find the maximum of $\alpha_e$ is
approx. $85\%$ of the predicted (constant) value of $\alpha_c$ in
$d=2$ and $\approx 70\%$ in $d=3$, instead of $\approx 90\%$ in
$d=4$. This refers to the prediction for $\alpha_c$ given the standard
values of $\kappa$, i.e.\ $0.3976\ldots$ in $d=3$ and $0.2$ in
$d=2$. The situation slightly improves if the value of the predicted
$\alpha_c$ is calculated instead with the $\kappa$ values actually
observed on the lattice. In $d=3$, this amounts to choosing
$\kappa=0.35$, see \Fig{fig:coup_forpaper}.

While the behavior of $\alpha_e$ in $d=2$ and $3$ is surprising and
not amenable to an immediate explanation, it is interesting to note
that the maximum of $\alpha_s(aq)$ shows a monotonic behavior from
$d=2$ to $d=4$, i.e., it is shifted to larger $aq$.  This follows
the general pattern observed also at finite $\beta$: the window in
which a scaling-like behavior is observed extends to much smaller
momenta in two dimensions than in three dimensions, before the
effective masses begin to dominate. In four dimensions, finally, the
window becomes so small that almost no scaling-like behavior can yet
be observed. The same mechanism seems here to shift the observed
maximum to ever smaller values of $aq$.

A possible explanation for these results would be that at large $aq$
discretization effects are visible while at small $aq$ a Gribov copy
effect causes the deviation of $\alpha_e$ from the continuum solution
($\alpha_e=\alpha_c=\text{const}$).

The second possibility motivates us to investigate the effect of the
$\max B$ gauge with a growing number of copies on $\alpha_e$. We have
used up to 600 Gribov copies per configuration on a $20^3$
lattice. The effect on the ghost propagator is strong, as is apparent
from \Fig{fig:ghdf_DIV_vs_axgf} (right) in the previous subsection. We
observe no `saturation'.

By virtue of \Eq{eq:def-alphas}, this effect induces an even stronger
change in the running coupling (see \Fig{fig:coup_varnaxgf}),
which is far from being balanced by a similarly strong change in the
gluon propagator. In fact, it was already stated in \cite{Maas:2009se}
that the effect of choosing the $\max B$ gauge on the gluon propagator
is relatively small for the employed lattice sizes.

Since the result overshoots the prediction this implies that it is
possible to select a value of $B$ such that actually an infrared
finite coupling could be generated. Such a choice leads to essentially
the same behavior, as is seen when artificially the number of copies
is restricted to $\approx 70$ in $d=3$. This is equivalent to the
statement that the locally extracted values of $\kappa$ and $\kappa_Z$
are consistent within errors in this region, see
\Fig{fig:averagekappa_glpetghp_axgf20vsnoaxgf_3d}.

There is currently no fully satisfactory explanation for the behavior
of the running coupling and the observed over-scaling. Even though a
$n_\text{copy}\to\infty$ extrapolation is not at all feasible from the
current data, the result in this limit on a $20^3$ lattice will be at
odds with the continuum prediction for $\alpha_c$. One possible
solution might be to consider larger lattices.

We have done similar simulations for the two-dimensional case, again
initially on a small lattice ($48^2$). The qualitative picture is
similar to the three-dimensional case, see \Fig{fig:coup_varnaxgf},
as a scaling region is also absent. As noted above, this is
surprising. In particular, when considering the scaling behavior
observed in two dimensions
\cite{Maas:2007uv,Cucchieri:2008fc,Cucchieri:2007rg,Pawlowski:2009iv} at volumes where
only very few Gribov copies are present
\cite{Maas:2008ri,Maas:2009se}.

Regarding the presence of possible discretization artifacts at large
momenta, one possible concern in this context is the lattice breaking
of rotational invariance at $\beta=0$. The fact that the dressing
functions on the lattice are invariant under the lattice isometries
does not imply that they are also invariant under the Euclidean $O(d)$
symmetry in $d$ dimensions. This implies that the dressing functions
depend on the orbits generated by the lattice isometries, but there
are different such orbits that correspond to the same $O(d)$ orbit
classified by the magnitude of momentum $q^2$ (see, e.g.,
\cite{deSoto:2007ht}).  These differences are 
of higher order in the lattice spacing and would vanish in the
continuum limit. There is no suppression of these differences in the
strong-coupling limit, however, and they will therefore show up in
differences of dressing functions evaluated along different momentum
directions. We assess these effects of the lattice breaking of
rotational invariance in Appendix \ref{app:momenta}. They are not as
dramatic as the Gribov effects, but the tendency of the fixed-point
couplings to decrease at large lattice momenta is within the
systematic uncertainty due to these effects.

\section{Conclusions}\label{sec:conclusion}

We close with a summary of our main results and a discussion of the
consequences.  In the present work we have qualitatively extended
previous studies on the strong-coupling limit $\beta=0$ of Landau
gauge SU(2) Yang--Mills theory on the lattice. In two and three
dimensions we have shown that the Gribov ambiguity is rather
strong. It especially affects the ghost propagator which neither
uniquely shows a scaling nor a decoupling behavior for small lattice
momenta. Indeed, subject to the number of gauge copies used, it even
shows over-scaling for small and intermediate momenta.  Our $\beta=0$
results also indicate that over-scaling can be achieved for any finite
lattice if enough gauge copies are taken into account. However, in the
continuum over-scaling is prohibited by the uniqueness proof put
forward in \cite{Fischer:2006vf,Fischer:2009tn,Huber:2007kc}. It is
not part of the one-parameter family of solutions
\cite{Fischer:2008uz,vonSmekal:2008ws} and might hint at the scaling
solution as the limit of the over-scaling lattice Landau gauge fixing.

In any case, the above results are in marked contrast to the results
in previous works where only a decoupling-type solution was obtained
at low momenta, see
\cite{Sternbeck:2008wg,Sternbeck:2008mv,Sternbeck:2008na,%
  Cucchieri:2009zt}. Evidently the interpretation of the $\beta=0$
results needs to be seriously reconsidered.

At large lattice momenta, the ghost propagator does not naively
exhibit scaling, as its local scaling coefficient, $\kappa$, rises
monotonically. This is neither a scaling behavior, nor is it a
decoupling-type behavior. Indeed, a logarithmic fit appears to be
possible, and has been related to similar fits for the decoupling
solution in the continuum \cite{Cucchieri:2009zt}. Note, however, that
this similarity is superficial because the prefactor of the
logarithmic term has the wrong sign. A naive extrapolation to finite
$\beta$ and the infinite-volume limit would lead to a pole at some
finite momentum. On the other hand, the gluon propagator shows
approximate scaling and hence a qualitatively different behavior from
the ghost. Such a different behavior would not be expected for a
decoupling-type solution.  A possible explanation for these surprising
results are strong but not unexpected discretization effects
associated with the unphysical $\beta\to0$ limit at large $aq$, see
Appendix \ref{app:momenta}.

Our results are best summarized in \Fig{fig:coup_varnaxgf} for the
running coupling. It is the only RG-invariant product of the two
dressing functions and is rather sensitive to the global aspects of
the gauge fixing. Its behavior at small lattice momenta confirms that
the Gribov ambiguity is much stronger than expected on the basis of
previous results. Extrapolating to finite $\beta$ in four dimensions,
we conclude that a resolution of the Gribov ambiguity requires far
more computational power than presently employed
\cite{Fischer:2007pf,Cucchieri:2007rg,Cucchieri:2008fc,%
  Sternbeck:2007ug,Bogolubsky:2009dc,Pawlowski:2009iv}. In particular this implies that
the present data cannot be used to exclude any subset of infrared
solutions. It is all within the uncertainties due to the Gribov
ambiguity. Indeed, we are confident that our data together with other
lattice and continuum results
\cite{Fischer:2008uz,vonSmekal:2008ws,PawlowskiSmekal,%
  Gomez:2009tj,Kondo:2009wk,Kondo:2009gc,Kondo:2009qz,%
  Kondo:2009ug,Maas:2008ri,Neuberger:1986xz,Sorella:2009vt,%
  vonSmekal:2007ns,vonSmekal:2008en,vonSmekal:2008es,%
  Zwanziger:2009je,Huber:2009tx,Binosi:2009qm,Boucaud:2008ji} can be
unified in an overall picture of Landau gauge Yang--Mills theory as put
forward in \cite{Fischer:2008uz,vonSmekal:2008ws,PawlowskiSmekal,Maas:2009se}. Its
further understanding would certainly provide  more insight
into the confining physics.

\section*{Acknowledgments}

We thank A.~Cucchieri, C.~S.~Fischer, T.~Mendes, M. M\"uller-Preussker and I.~Stamatescu for discussions
and comments on the manuscript. This work is supported by the
Helmholtz Alliance HA216/EMMI and the Helmholtz International Center
for FAIR within the LOEWE program of the State of Hessen.  A.\ M.\ was
supported by the FWF under grant number M1099-N16, and A.~S. by the
Australian Research Council and the SFB/TR~55. D.~S. acknowledges support by the
Landesgraduiertenf\"orderung Baden-W\"urttemberg via the Research
Training Group ``Simulational Methods in Physics''. The numerical
simulations were carried out on bwGRiD (http://www.bw-grid.de), member
of the German D-Grid initiative, on the compute cluster of the ITP,
University of Heidelberg, and partly at the HPC cluster of the
University of Graz.

\appendix

\section{Plane-wave vs.\ point source}\label{app:source}

\begin{figure}[t]
\begin{center}
  \includegraphics[clip,width=0.92\columnwidth]{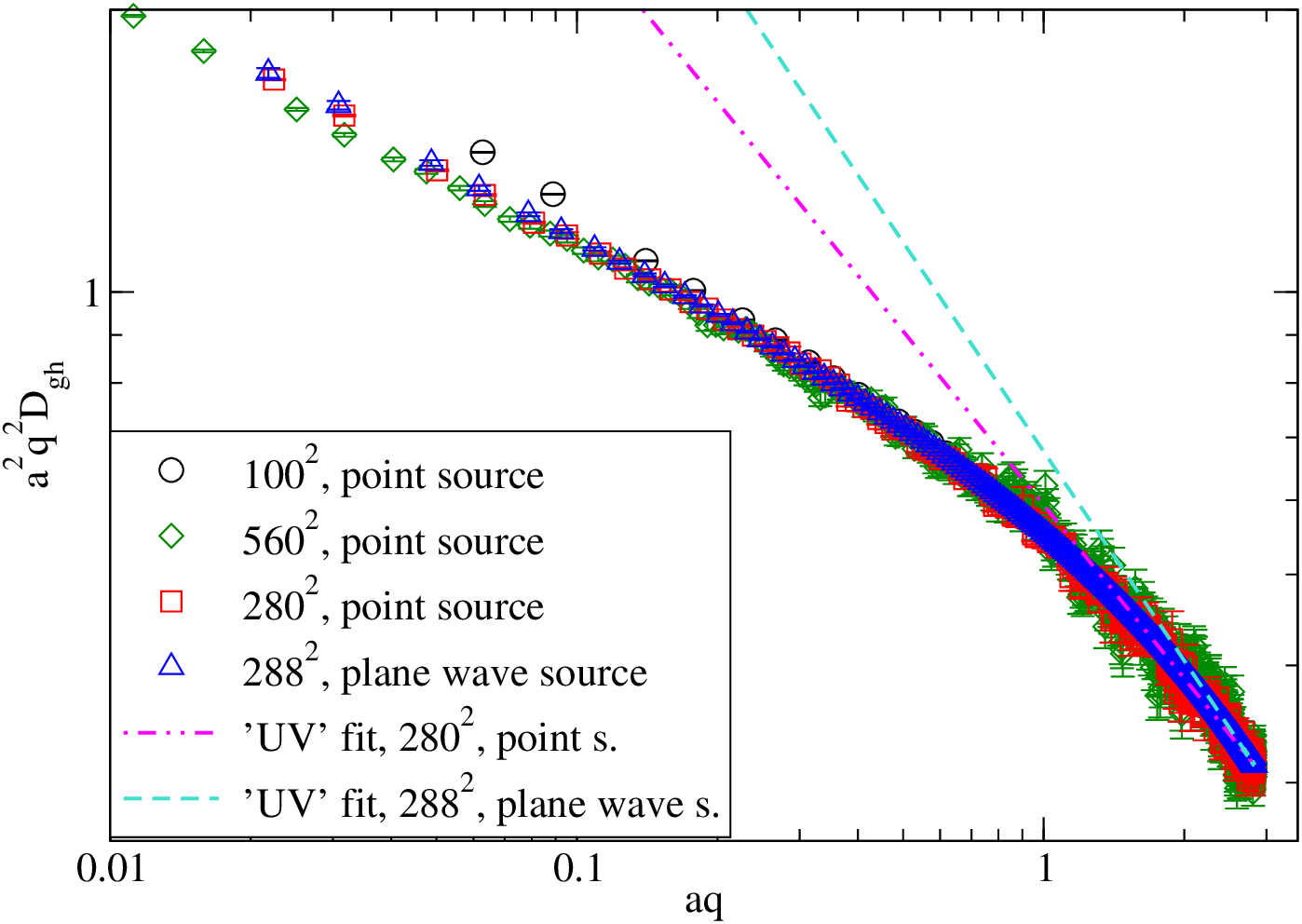}
  \caption{\label{fig:app:ghdf_beta0_varls_2d_forpaper}Ghost dressing
    function in $d=2$. In order to draw conclusions about the behavior
    at large lattice momenta, the plane-wave source method is clearly
    preferable to the point-source method: only 76 meas.\ on $288^2$
    (plane-wave source) vs.\ $\approx 37000$ on $280^2$ (point source)
    yield much less fluctuations.}
\end{center}
\end{figure}

\begin{figure}[t]
\begin{center}
  \includegraphics[clip,width=0.92\columnwidth]{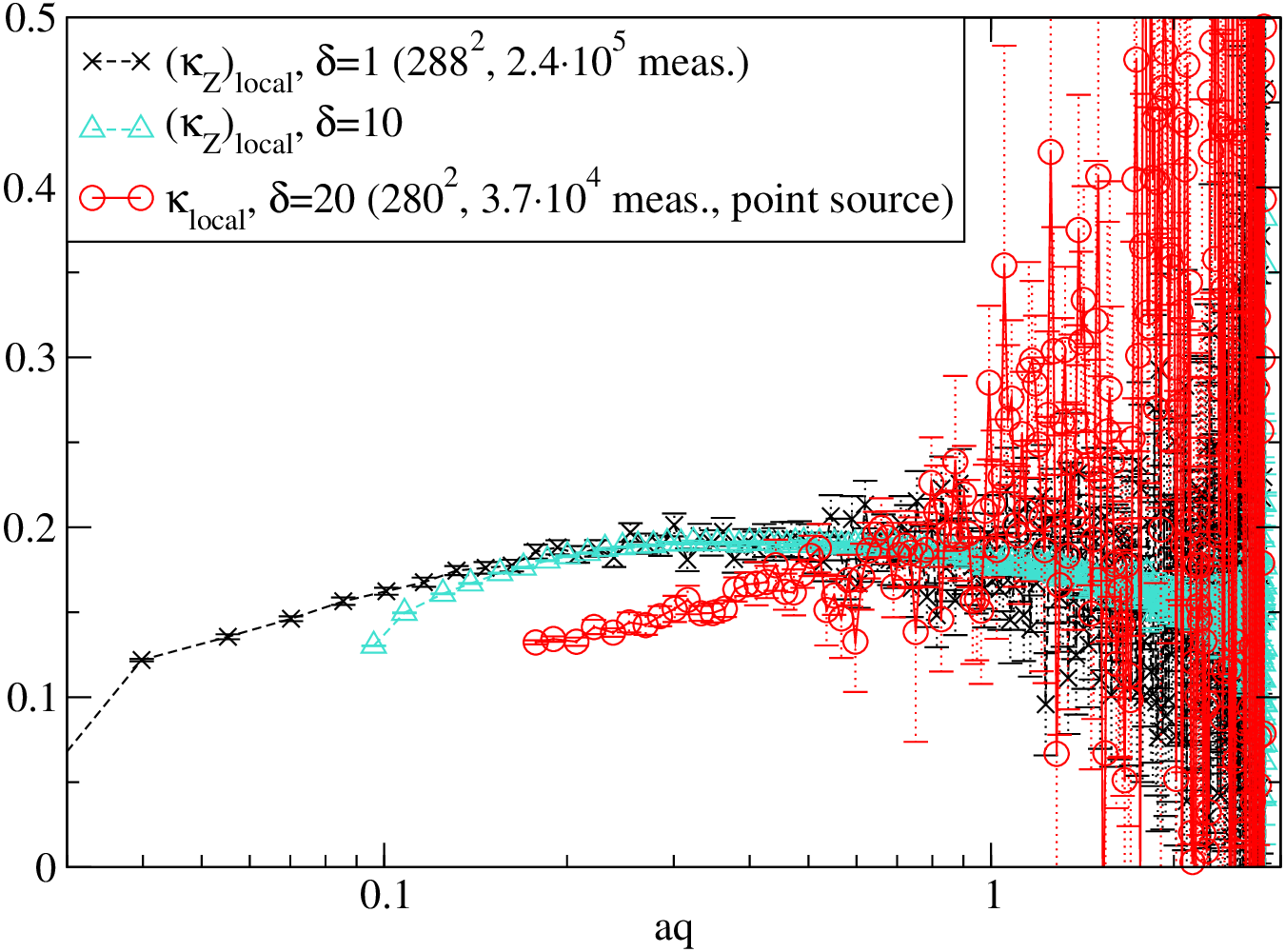}
  \caption{\label{fig:app:averagekappa_glpetghp_ls280resp288_2d_varwl-point}`Local
    $\kappa$' for a point source from 37000 measurements and
    calculated from (cylinder cut) momentum norms that are $\delta=20$
    indices apart. Compare to the result from a plane-wave source with
    76 meas. at $V=288^2$ in
    \Fig{fig:averagekappa_glpetghp_ls288_2d_varwl}.}
\end{center}
\end{figure}

We expand here the study of \cite{Cucchieri:2006tf}, which showed that
the method to obtain the inverse of the Faddeev--Popov operator to
calculate the ghost propagator has quite an impact on the statistical
accuracy of the calculation.  Especially, for extracting information
about the local behavior it is advantageous to employ a plane-wave
source \cite{Cucchieri:1997dx} instead of a point source
\cite{Boucaud:2005gg} for the conjugate gradient method used to invert the
operator. This means that the Faddeev--Popov operator is inverted on a
vector of plane waves, $s_b^{a,x}(k)=\delta^{ab}e^{ik\cdot x}$,
instead of a point source, $s_b^{a,x}=\delta^{ab}(\delta_{x,0}-1/V)$,
with $k\cdot x=2\pi\sum_\mu k_\mu x_\mu /L_\mu$.

The contrast is illustrated in
\Fig{fig:app:ghdf_beta0_varls_2d_forpaper}, which already suggests
that the point source makes a power-law fit at large lattice momenta,
and thereby the identification of a scaling region with an unique
exponent $\kappa$, considerably more difficult. This statement can be
made more precise, e.g., by performing power-law fits at large lattice
momenta and comparing the respective values of $\chi^2/\ndf$. The
reason is that a plane-wave source averages over all possible
inversion points, and therefore provides a volume-factor less noise
than a point source. The drawback of a plane-wave source is that an
inversion is required for every momentum value.

This problem is further illustrated in
\Fig{fig:app:averagekappa_glpetghp_ls280resp288_2d_varwl-point} for a
locally defined $\kappa$, in contrast to
\Fig{fig:averagekappa_glpetghp_ls288_2d_varwl} (plane-wave source). Of
course, in the limit of infinite statistics, both methods will yield
the same result. Using a point source to determine the value of
$\kappa$ at large momenta, a value of $0.31$ would be obtained.

\begin{figure}[t]
\begin{center}
\includegraphics[clip,width=0.92\columnwidth]{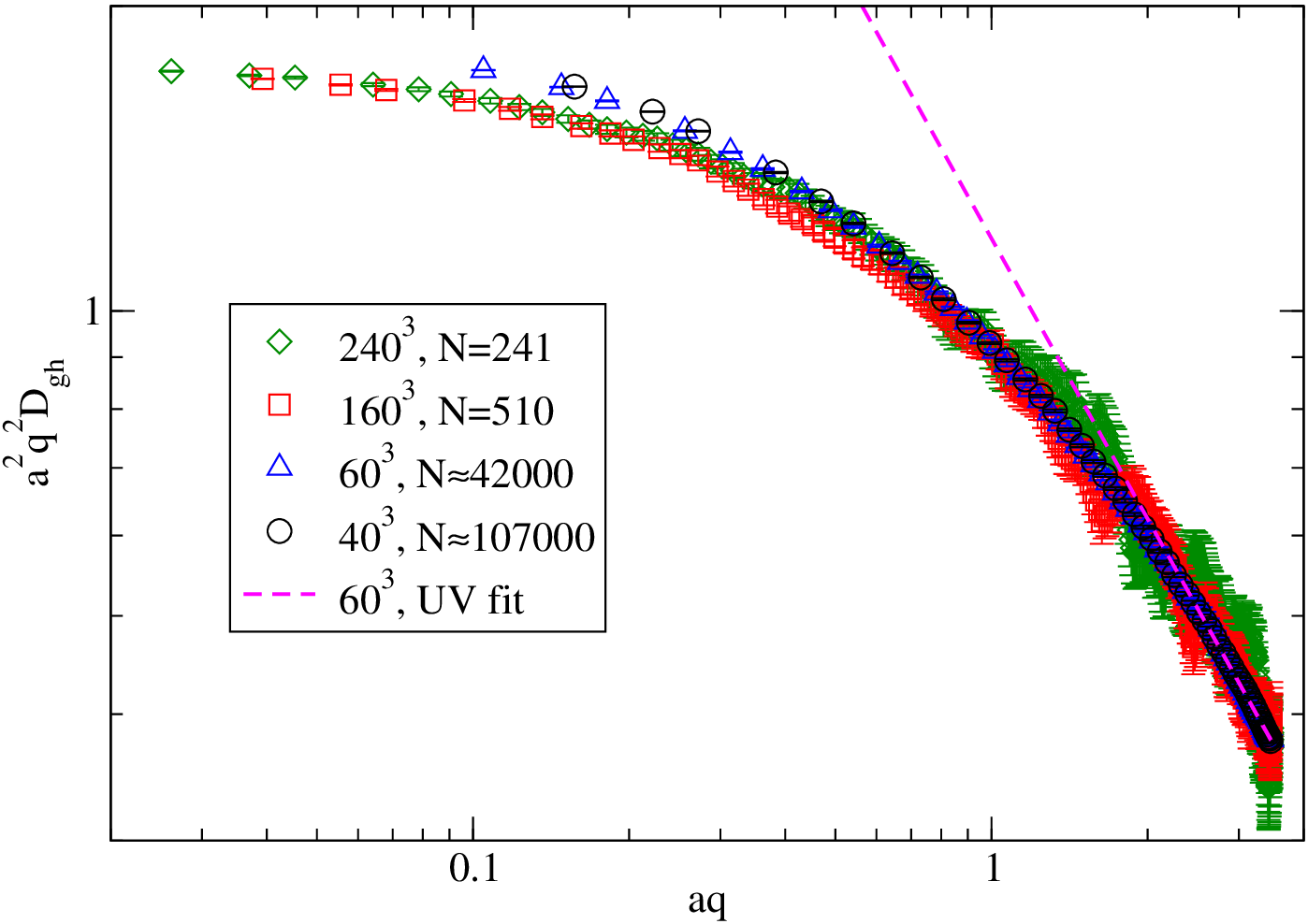}
\caption{\label{fig:ghdf_beta0_varl}The ghost dressing function in
  three dimensions using the point-source method. The numbers $N$ of
  independent configurations are given in the legend. For an
  explanation of the wiggles at large momenta see
  \cite{Cucchieri:2006tf}.}
\end{center}
\end{figure}

\begin{figure}[t]
\begin{center}
  \includegraphics[clip,width=0.92\columnwidth]{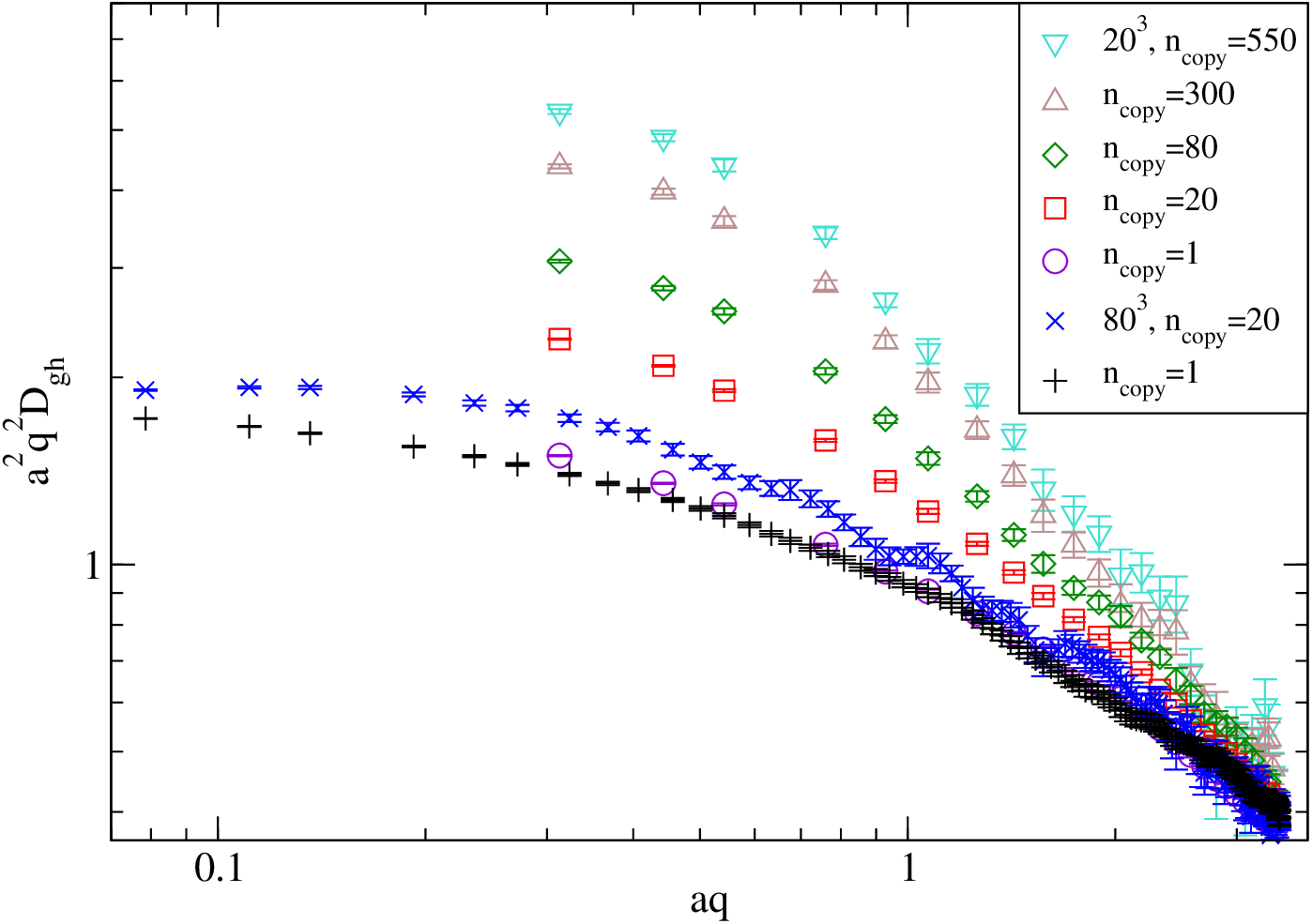}
  \caption{\label{fig:ghdf_axgf_nrcopy20_beta0_ls20et80_3d} Finite-volume effect for minimal Landau gauge and for $\max B$ gauge with
    various numbers of copies. Point-source method.}
\end{center}
\end{figure}

Similar considerations apply in all dimensions. The result for the
point-source method in three dimensions is shown in
\Fig{fig:ghdf_beta0_varl}. The fluctuations are greatly enhanced
compared to the plane-wave result shown in
\Fig{fig:ghdf_beta0_plane_64_withfit}. Even with
$N\in\mathcal{O}(10^5)$ configurations, they are still larger than in
the latter case. Correspondingly, the region where a
power-law fit at large $aq$ with $\chi^2/\ndf<1$ is
possible is much smaller.

There is also a note in conjunction with the $\max B$ gauge
employed. The ghost propagator using the point source exhibits much
larger statistical fluctuations. Consequently, when triggering on
large fluctuations, as when implementing the $\max B$ gauge, it is
likely that in a sample of finite statistics the ghost propagator will
be larger in the $\max B$ gauge than when using a plane-wave
source. Due to its inherent averaging over the lattice the plane-wave
source method smooths these exceptional fluctuations in many more
cases. Only when volume-times more statistics have been used for the
point-source method than for the plane-wave source method, a similar
smoothing effect can be expected to take place. Again, in the limit of
infinite statistics both methods will then yield the same result. A
measure of this, aside from the ghost propagator directly, is the
comparison of average vs.\ median. At the present statistics, in both
cases both measures significantly deviate but the difference is much
larger for the point-source method than for the plane-wave source
method. Also, when using the median instead of the average the results
for both methods are more similar.

As an example, the results at finite statistics for the point-source
method are shown in
\Fig{fig:ghdf_axgf_nrcopy20_beta0_ls20et80_3d}. Comparing to
\Fig{fig:ghdf_beta0_plane_noaxgfvsaxgf20_varl} from the main text
displays exactly this point.

\begin{figure}[t]
\begin{center}
\includegraphics[clip,width=0.92\columnwidth]{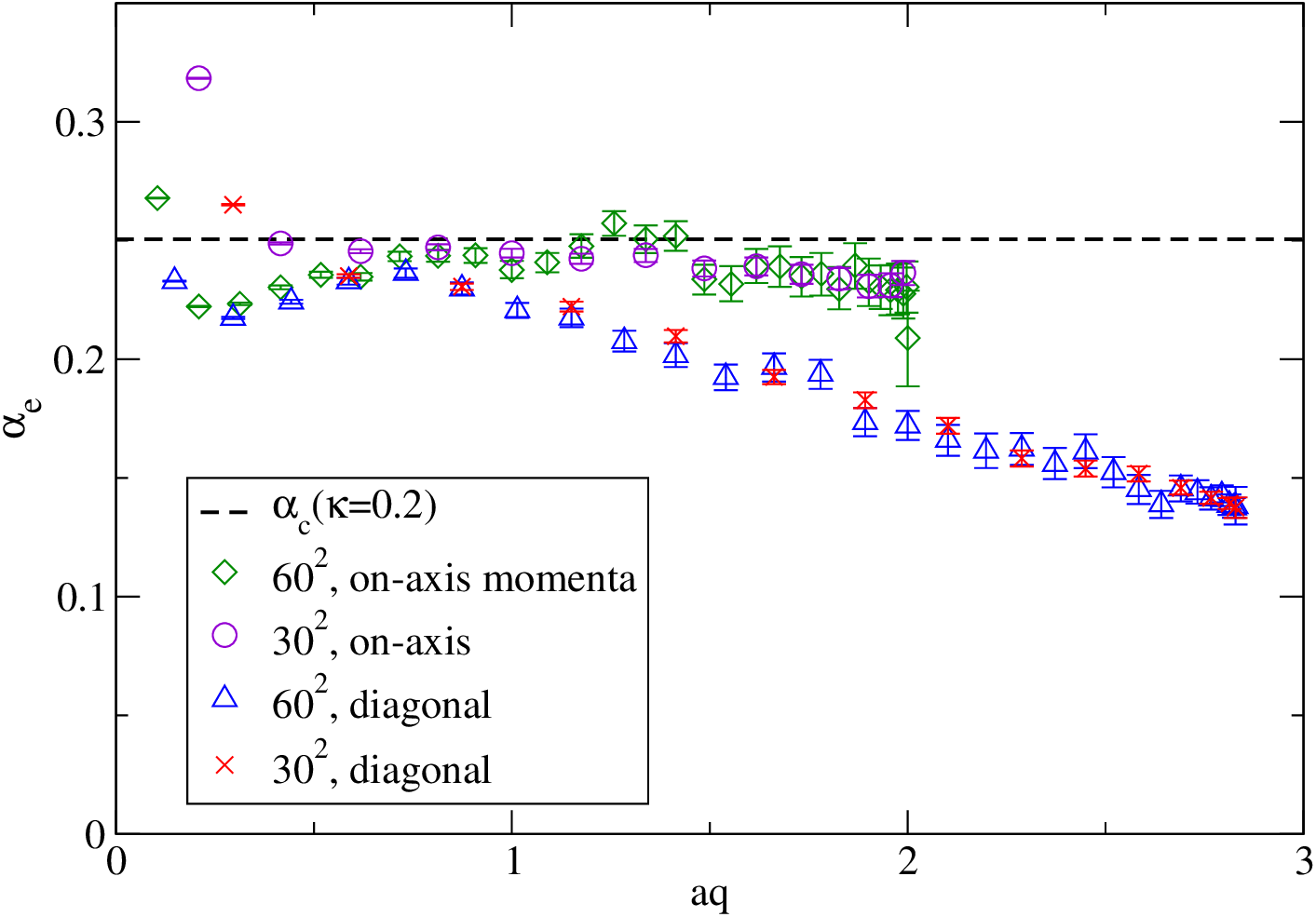}
\caption{\label{fig:coup-2d-onaxis-diag}Effective running coupling
  in $d=2$ for two different choices of lattice momenta.}

\end{center}
\end{figure}

\section{Discretization effects}\label{app:momenta}

\begin{figure}[t]
\begin{center}

\vspace*{.8cm}
\includegraphics[clip,width=0.92\columnwidth]{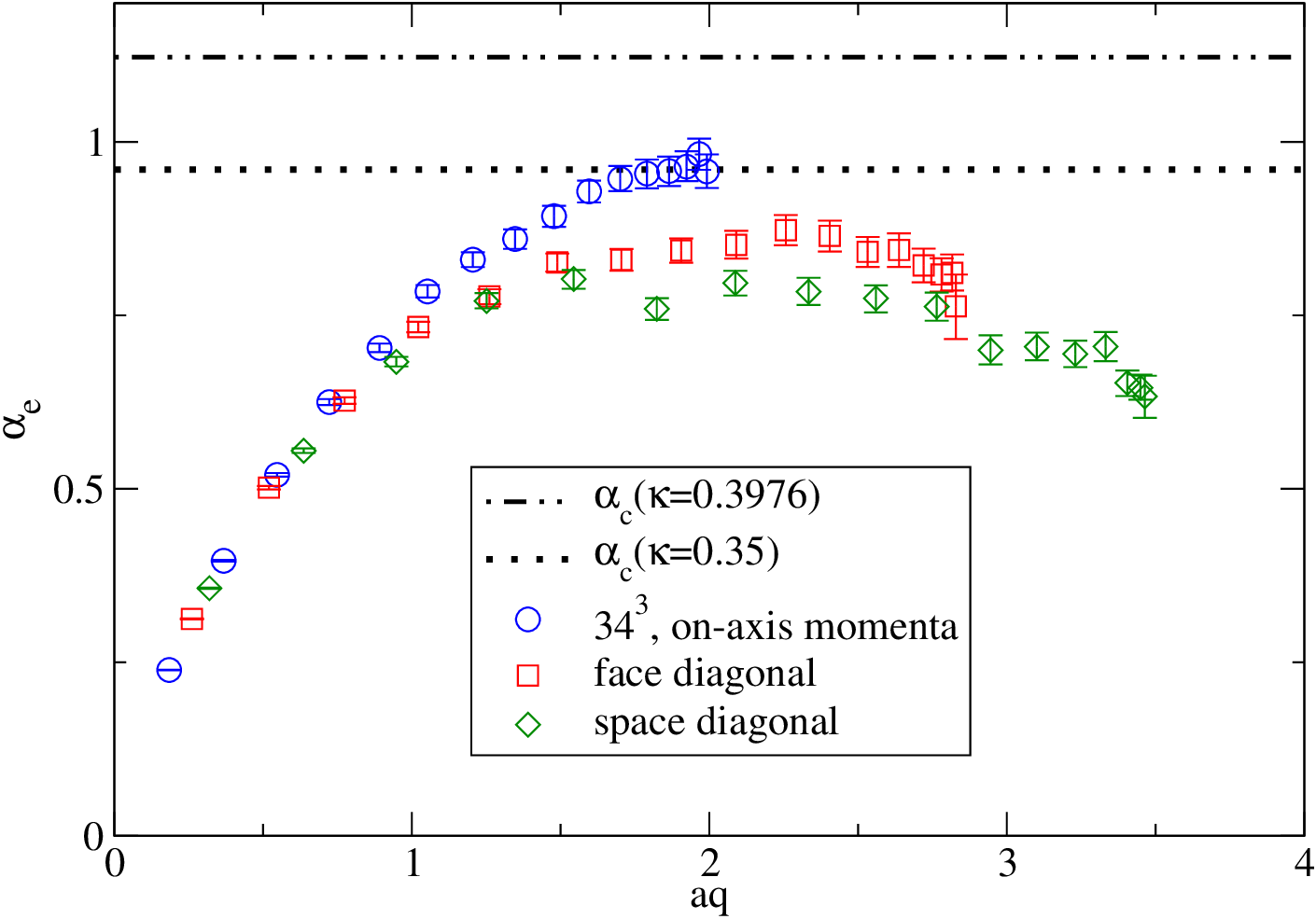}
\caption{\label{fig:coup-3d-onaxis-facediag-spacediag}Effective
  running coupling in $d=3$ for three different choices of lattice
  momenta.}

\end{center}
\end{figure}

As noted above, we have usually applied a cylinder cut to the
momenta. In order to obtain a rough estimate of possible
discretization artifacts, we have also calculated the running coupling
for the opposite case of on-axis momenta, $k_i=k\delta_{ij}$ for all
fixed $j$. Interestingly, the effective running coupling thus obtained
is compatible with a constant value at large lattice momenta close to
the continuum prediction \cite{Lerche:2002ep}, see
\Fig{fig:coup-2d-onaxis-diag}. These data have been produced with the
point-source method, which suffices to clearly observe the qualitative
effect. Data at diagonal momenta, which form a subset of the momenta
surviving the cylinder cut, are shown for comparison. In $d=3$, the
difference between the choices of momenta is again clearly visible,
see \Fig{fig:coup-3d-onaxis-facediag-spacediag}, though less
pronounced than in the lower-dimensional case. This suggests an even
weaker effect for $d=4$, where indeed the running coupling has been
found to tend towards a constant value at large $aq$
\cite{Sternbeck:2008wg,Sternbeck:2008mv}. These early results
underline the serious discretization problem faced at large $aq$ at
$\beta=0$ and call for further investigations along these lines,
e.g. in $\max B$ gauge.

\bibliographystyle{apsrev-SLAC}
\bibliography{references}

\end{document}